\pdfoutput=1
\documentclass{emulateapj}

\slugcomment{Submitted to ApJ}

\begin{document}
%------------------------------------------------------------------------------

\title{Application of the Cubed-Sphere Grid to Tilted Black-Hole Accretion Disks}

\author{P. Chris Fragile and Christopher C. Lindner \altaffilmark{1}}
\affil{Department of Physics and Astronomy, College of Charleston,
Charleston, SC 29424}
\email{fragilep@cofc.edu}

\and

\author{Peter Anninos and Jay D. Salmonson}
\affil{Lawrence Livermore National Laboratory, Livermore CA 94550}

\altaffiltext{1}{Astronomy Department, University of Texas, 1 University Station, C1400, Austin, TX 78712}
\date{{\small    \today}}
\date{{\small   \LaTeX-ed \today}}
%-----------------------------------------------------------------------------

\begin{abstract}
In recent work we presented the first results of global general relativistic magnetohydrodynamic (GRMHD) simulations of tilted (or misaligned) accretion disks around rotating black holes. The simulated tilted disks showed dramatic differences from comparable untilted disks, such as asymmetrical accretion onto the hole through opposing ``plunging streams'' and global precession of the disk powered by a torque provided by the black hole. However, those simulations used a traditional spherical-polar grid that was purposefully underresolved along the pole, which prevented us from assessing the behavior of any jets that may have been associated with the tilted disks. To address this shortcoming we have added a block-structured ``cubed-sphere'' grid option to the Cosmos++ GRMHD code, which will allow us to simultaneously resolve the disk and polar regions. Here we present our implementation of this grid and the results of a small suite of validation tests intended to demonstrate that the new grid performs as expected. The most important test in this work is a comparison of identical tilted disks, one evolved using our spherical-polar grid and the other with the cubed-sphere grid. We also demonstrate an interesting dependence of the early-time evolution of our disks on their orientation with respect to the grid alignment. This dependence arises from the differing treatment of current sheets within the disks, especially whether they are aligned with symmetry planes of the grid or not.
\end{abstract}


\section{Introduction}
\label{sec:intro}

We have recently undertaken a series of numerical studies of titled
accretion disks around rapidly rotating black holes, first in the
hydrodynamic \citep{fra05} and then in the magnetohydrodynamic (MHD)
\citep{fra07b} limits. All of these simulations have been fully
general relativistic, using the Kerr-Schild metric to represent the
spacetime of the black hole.

Tilted accretion disks are of particular interest because they are
subject to differential warping due to the Lense-Thirring precession
of the rotating black hole. For very thin disks, close to the black hole
the competition between the differential twisting and ``viscous''
damping causes the angular momenta of the disk and hole to align.
Further out in the disk, beyond some warp radius, the disk 
maintains its misaligned state.

For moderately thin to thick disks, such as the ones we simulated previously, the
situation is more complex and interesting. The primary difference is that warping is transported via bending waves rather than diffusively, as for thin disks. One consequence of this is that the midplane of a thick disk does not tend to align with the symmetry plane of the black hole at small radii, as in the thin disk case. In fact, the relative tilt between the black-hole and disk angular momenta can {\em increase} at small radii. Having the tilted disk penetrate very close to the black-hole has many interesting consequences. For instance, we found
that accretion onto the hole occurs predominantly through two
opposing ``plunging streams'' that start from high latitudes with
respect to both the black-hole and disk midplanes \citep{fra07a}.
There is also a strong epicyclic driving within the disk
attributable to the gravitomagnetic torque of the misaligned
(tilted) black hole \citep{fra08b}. The induced motion of the gas can be
coherent over the scale of the entire disk. The gas also experiences
periodic (twice per orbit) compressions. The compressions occur as
the gas orbits past the line-of-nodes between the black-hole
symmetry plane and disk midplane. Near the black hole these
compressive motions can become supersonic and transform into a pair of
quasi-stationary shocks. The shocks act to
enhance angular momentum transport and dissipation near the hole,
forcing some material to plunge toward the black hole from well
outside the innermost stable circular orbit. Finally, because we are
simulating disks with finite radial extents and fast sound-crossing
times, the torque of the black hole causes the entire disk body to
precess globally.

The main shortcoming of our work so far comes from limitations
imposed on us by our use of a spherical-polar grid. First,
construction of a uniform spherical polar grid in three-dimensions
results in very small zones surrounding the two poles, where all of
the lines of longitude converge. These very small zones constrain
the Courant-limited timestep to be exceedingly small, such that the
required CPU cycle count becomes prohibitively large. To avoid this
problem, researchers have either excised a small conical section around
each pole \citep[e.g.][]{dev03a} or used a lower grid resolution near the
poles \citep{fra07b}. Although these techniques are reasonable when
one is primarily interested in studying the equatorial region (where
a disk may form), these are not satisfactory when one is interested in what is happening in the polar regions
(where jets may form).  A second problem with the spherical-polar
grid is that the poles themselves actually represent coordinate singularities,
which present significant challenges for numerical advection
and curvature coupling schemes (e.g. solving Riemann curvature source terms).

For these reasons we have added the cubed-sphere grid \citep{sad72, ron96}
as an option within our numerical code, Cosmos++.  The advantage of
this grid construction is that its topological properties more
closely resemble a Cartesian coordinate system than a spherical-polar
system. The cubed-sphere grid uses a more uniform zone spacing than spherical
polar, so the timestep can remain reasonably large even in high
resolution simulations. Also important, the grid does not contain
any coordinate singularities except at the origin, which is not a
concern for our intended use since we truncate the grid just inside the event
horizon of the black hole. Ours is not the first application of the
cubed-sphere grid to problems in computational astrophysics; it has
been used previously to study accretion onto rotating stars with inclined magnetic fields \citep{kol02,rom03} and a few problems in stellar evolution \citep{dea05,dea06}. However this is the
first application of this grid to the study of black-hole accretion
disks and their attendant jets.

The paper is organized as follows: \S \ref{sec:cubedsphere} describes the cubed-sphere mesh
in detail and our particular implementation. In \S \ref{sec:gradtest} we
discuss results of basic gradient tests on the cubed-sphere grid.
In \S \ref{sec:disks} we compare two sets of
numerical simulations of black-hole accretion disks. In the first set we
compare simulations of disks accreting onto a
Schwarzschild black hole. We compare different grids, resolutions, and orientations of the disk with respect to the grid.
Because these simulations use a Schwarzschild black hole, the orientation should have no physical meaning. However, we show that there
are, nevertheless, considerable differences in their evolution at early times. We
present the case that the differences have to do with the differing
treatments of the midplane current sheets in the disks, which forms from the differential winding of our initial poloidal field loops. Finally we
compare simulations of tilted accretion disks around Kerr black
holes. We use a tilt of $\beta_0=15^\circ$ and a spin of $a/M_\mathrm{BH}=J_\mathrm{BH}/M^2_\mathrm{BH}=0.5$,
in geometrized units where $G=c=1$, and $M_\mathrm{BH}$ and $J_\mathrm{BH}$ 
are the mass and angular momentum of the black hole, respectively. 
One of these simulations is run
on a spherical-polar mesh, the others on the cubed-sphere grid. We
demonstrate that, for the most part, the simulations agree very well.

\section{The Cubed Sphere}
\label{sec:cubedsphere}

The cubed-sphere grid gets its name from its construction -- it is
actually composed of six ``blocks'' that are morphed into segments of
a sphere. Each block is constructed of segments of concentric radial shells. In
the present work, these shells are spaced exponentially based upon
their distance from the hole, similar to a logarithmic radial
coordinate. The other two coordinates are constructed such that, on
any given block, the grid lines trace out ``great circles'' on each
radial shell segment. It is as if there are two longitude coordinates,
$\phi_1$ and $\phi_2$, on each block. The range of $\phi_1$ and
$\phi_2$ on each block is $\pi/2$ so that the full $4\pi$ steradian
is covered by the six blocks.

The difficulty with the cubed-sphere grid is that the ``great
circles'' cannot be made continuous across all six blocks, and hence the
block-structured nature of the mesh. Stated
differently, the coordinates $\phi_1$ and $\phi_2$ cannot maintain a
consistent orientation across all blocks. At each block boundary, the
coordinate system has a discontinuous jump. Fortunately this can be
handled with the proper application of boundary conditions and
communication between blocks,
%as well as limiters within the advection routine, 
as we shall describe.

Another problem with the cubed-sphere grid is that the $\phi_1$ and
$\phi_2$ coordinates are not orthogonal. There are
techniques available to try to improve the orthogonality of the
cubed-sphere grid at the cost of reducing its uniformity. However, such
techniques have been shown not to perform significantly better than
the standard cubed-sphere grid implemented here \citep{put07}. 
Furthermore, such techniques are not necessary
in our Cosmos++ code, which is designed with tremendous mesh
flexibility to handle a variety of grids including fully unstructured
and non-orthogonal ones.

\subsection{Implementation within Cosmos++}

When working with more traditional spherical-polar meshes, 
the Cosmos++ code actually
evolves the MHD equations in a generalized coordinate system
$\{x_0,x_1,x_2,x_3\}$, with the curvature implemented through
metric terms. This is done even for the Newtonian formulation. 
The corresponding physical coordinates in general relativity are the
Kerr-Schild polar coordinates $\{t,r,\theta,\phi\}$. For the
cubed-sphere, instead, we construct the grid in physical space using the
Kerr-Schild Cartesian coordinate system $\{t,x,y,z\}$. The two Kerr-Schild coordinate systems
are related through the following transformations:
\begin{eqnarray}
x & = & r \sin \theta \cos \phi - a \sin \theta \sin \phi ~, \nonumber
\\
y & = & r \sin \theta \sin \phi + a \sin \theta \cos \phi ~, \nonumber
\\
z & = & r \cos \theta ~,
\label{eqn:KStransform}
\end{eqnarray}
or
\begin{eqnarray}
r^2 & = & \frac{ (x^2+y^2+z^2-a^2) + \sqrt{(x^2+y^2+z^2-a^2)^2 + 4a^2
z^2}}{2} ~, \nonumber \\
\sin \theta & = & \left( \frac{x^2+y^2}{r^2+a^2} \right)^{1/2} ~,
\nonumber \\
\cos \theta & = & \frac{z}{r} ~, \nonumber \\
\sin \phi & = & \frac{ry-ax}{\sqrt{(r^2+a^2)(x^2+y^2)}} ~, \nonumber
\\
\cos \phi & = & \frac{rx+ay}{\sqrt{(r^2+a^2)(x^2+y^2)}} ~.
\end{eqnarray}

\begin{figure*}
%\plotone{Cubed_Sphere_blocks.eps} 
\plotone{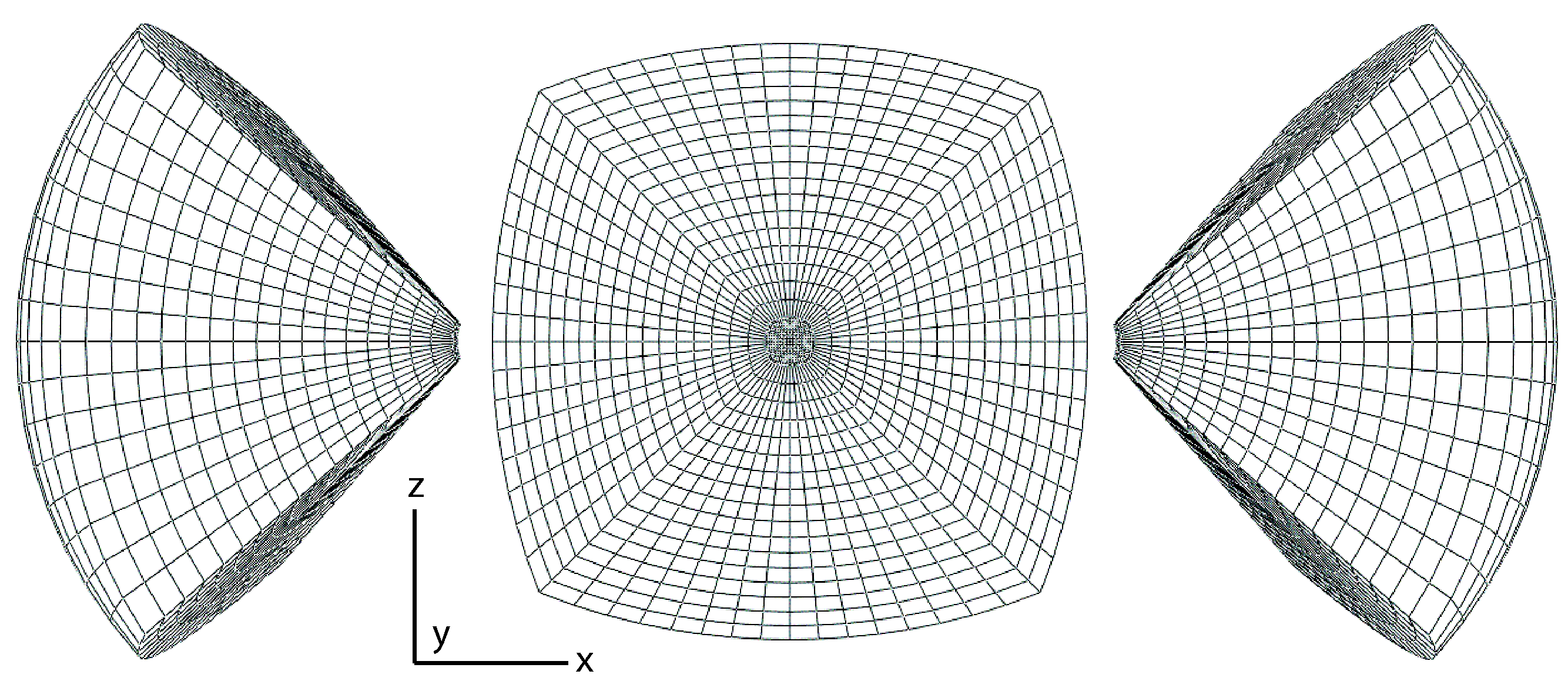}
\caption{Examples of blocks 0 ({\em right}), 1 ({\em center}), and 2 ({\em left}) that might make up a cubed-sphere grid. Note that in this illustration we use a very low resolution for clarity.
\label{fig:cubes}}
\end{figure*}

Ultimately the Cosmos++ code just needs to know the coordinate locations 
of all the zone vertices. From those it is able to fully reconstruct 
all of the necessary zone properties such as volumes and face areas. 
We find it easiest for the cubed-sphere grid to start from the cubed-sphere
coordinates $\{r, \phi_1, \phi_2\}$ of each vertex, then use the transformations given in Appendix \ref{app:transform} to convert to the Kerr-Schild polar 
coordinates $\{r,\theta,\phi\}$, and finally use equation (\ref{eqn:KStransform}) to obtain the correct Kerr-Schild Cartesian coordinates $\{x,y,z\}$ from the polar ones.
For convenience we label the six blocks 0-5, with their orientations described in Appendix \ref{app:transform}. Samples of blocks 0, 1, and 2 are illustrated in Figure \ref{fig:cubes}.

%\clearpage
%\clearpage

One consequence of using Kerr-Schild Cartesian coordinates is that,
whereas in the spherical-polar case we were able to tilt the black
hole with respect to the grid \citep{fra05,fra07a,fra07b}, in the
cubed-sphere case the black hole must remain aligned with the grid
for a rotating black hole ($a \ne 0$). This is because in
Kerr-Schild Cartesian coordinates, the $z$-axis is chosen to be the
spin axis of the black hole and the event horizon is only symmetric
about this axis. Thus, in order to get the inner boundary of the
cubed-sphere grid to align with the black-hole event horizon, the black-hole spin axis must align with the grid $z$-axis.

\subsection{Boundary Conditions}
As we said, one of the difficulties with the cubed sphere is that
the coordinates are not continuous across block boundaries. This
requires some care when setting up communications between blocks.
Even once the communication pattern between blocks is established,
there are certain subtleties about the grid that must be dealt with.
For instance, as we show in \S \ref{sec:gradtest}, the gradient
operator can only be made to converge properly (second order) if a set of ghost
zones are constructed that are an extension of the coordinates on
the current block. However, such ghost zones then do not correspond
directly to any of the zones on the neighboring block; instead they
tend to straddle more than one zone, and a simple domain exchange
is not exactly valid. Fortunately, at any inter-block
boundary it is only one of the $\phi$ coordinates that is
discontinuous; the radial coordinate and the other $\phi$ coordinate
are consistent across any given inter-block boundary \citep{ron96}. Therefore, for a single layer grid with uniform zone spacing, the
ghost zones of one block will never overlap more than two zones on
the neighboring block. 
In such a case, we can get away with applying a
boundary condition that simply fills the ghost zone with a field
$F_W$ that is a weighted average of the fields in the two real zones
it overlaps, $F_0$ and $F_1$. The weighted average we use is
\begin{equation}
F_W = \frac{F_0(L-\vert \mathbf{x}_0 - \mathbf{x} \vert) +
F_1(L-\vert \mathbf{x}_1 - \mathbf{x} \vert)}{L} ~,
\label{eqn:fw}
\end{equation}
where
\begin{equation}
L = \vert \mathbf{x}_1-\mathbf{x}\vert +
\vert\mathbf{x}_0-\mathbf{x}\vert ~,
\end{equation}
and $\mathbf{x}$, $\mathbf{x}_0$, and $\mathbf{x}_1$ are the
coordinate centers of the ghost zone and the two real zones it
overlaps on the neighboring block, respectively. This weighting
scheme is applied any time a normal domain exchange would be needed
between neighboring processors, such as after fields are updated,
but before any gradients are taken. Something slightly different
must be done for advection as explained in the next section.

\subsection{Advection}

Because Cosmos++ was written using finite volume methods, and 
designed for arbitrary mesh topologies, few changes were needed to
apply the code to the cubed-sphere mesh. The one thing (in addition to
the ghost zone construction) that was modified, 
if only slightly, was the algorithm for advection. A number of 
different options for advection are available in Cosmos++ including 
upwind subzonal polyhedral reconstruction and global monotonic flux 
limiter methods, described in \citet{ann05}. These methods are designed 
to operate on multi-dimensional vector quantities (e.g., gradients) 
constructed from the convex attributes of arbitrary covariant cell 
geometries and connectivities. However, for the cubed-sphere mesh we 
found that flux estimates performed with local one-dimensional limited 
projections (or differences) across individual cell faces are generally 
more robust than computing vector fluxes across the entire cell structure, 
even with appropriate multi-dimensional flux limiters. 

The method is only slightly modified from that presented in \citet{ann05}, so
we present only an abbreviated discussion. The advection terms are solved for each evolved field 
quantity using an upwind time-explicit, first order forward Euler scheme 
with appropriately time-centered fluxes. Letting ${\bf F}$ represent any 
of the evolved fields (or their consistent transport counterparts 
with ${\bf F} \rightarrow {\bf F}/D$, where $D$ is the mass density), 
the discrete finite-volume representation of the advection source 
term can be written
\begin{equation}
\partial_i ({\bf F} V^i) = -\frac{1}{V_z} \sum\limits^\mathrm{faces}_f 
           ({\bf F}^*~V^i~A_i)_f  ~,\label{eqn:fv_adv}
\end{equation}
where $V_z$ is the local donor cell volume of zone $z$, $(A_i)_f$ is the 
inward pointing area normal vector associated with face $f$ of the donor 
cell, and $(V^i)_f$ is the face-centered velocity defined as a weighted 
average across neighboring cells. In \citet{ann05}, the quantity 
$({\bf F}^*)_f$ represents piecewise linearly reconstructed zone-centered 
fields extrapolated to each cell face by a monotonic Taylor's series expansion,
${\bf F}^* = {\bf F}_{z} + (\partial_i {\bf F})^L_{z} (r^i - r^i_{z})$, 
projected from the donor cell center $r^i_{z}$ to either the face center 
$r^i = r^i_f$ or the advection control volume center 
$r^i = r^i_f - (\Delta t/2) (V^i)_f$, over a time-step interval $\Delta t$. 
The zone-centered limited gradient $(\partial_i {\bf F})^L_z$
forces monotonicity in the extrapolated fields using polyhedral subzonal 
interpolations and control volume integrals to construct upwind, downwind
and centered variations. 
The difference here, for the cubed-sphere, is that the monotonic 
multi-dimensional gradient is replaced by a local one-dimensional 
calculation separately across each donor cell face, and along the direction 
of the cell face normal (perpendicular to the cell face) using the 
generalized minmod limiter in the form
\begin{equation}
\nabla {\bf F}_{\perp} = \left[      \frac12\left(\frac{a}{|a|} + \frac{b}{|b|}\right)  
\times \frac12\left(\frac{a}{|a|} + \frac{c}{|c|}\right) 
\times \frac12\left(\frac{b}{|b|} + \frac{c}{|c|}\right) \right] \min\left(|a|, ~|b|, ~|c|\right) ~,
\end{equation}
where $a=(1+\lambda) \nabla {\bf F}_{D}$, $b=\nabla {\bf F}_{C}$, 
$c=(1+\lambda) \nabla {\bf F}_{U}$, $\lambda$ is an order parameter between 
zero and unity specifying the steepness of the applied limiter, and 
$\nabla {\bf F}_{U}$, $\nabla {\bf F}_{D}$, and $\nabla {\bf F}_{C}$ are 
the upwind, downwind, and center-difference gradients, respectively. 
The upwind and downwind gradients are defined as 
$\nabla {\bf F}_{U(D)} = k~\delta {\bf F}/\delta s$, where $k = \pm 1$ 
depending on the upwind direction with respect to the coordinate orientation, 
$\delta {\bf F} = {\bf F}_z - {\bf F}_{opp}$ is the difference between donor 
and opposite cell field values, and $\delta s$ is the magnitude of the 
distance between donor and neighbor cell centers. The center-difference 
representation of the gradient is approximated as 
$\nabla {\bf F}_C = \sum_{faces} (k/2)(\delta {\bf F}/\delta s)$, 
where the sum is over opposite cell face pairs. A projected estimate for 
the advected fields contributing to the flux in equation (\ref{eqn:fv_adv}) 
at each cell face is provided by the donor cell as 
${\bf F}^* = {\bf F}_z + \delta {\bf F} = {\bf F}_z - k~\nabla {\bf F}_{\perp} \times \delta r$, 
where $\delta r = |\vec{x}_f - \vec{x}_z - 0.5\Delta t (V^i A_i) \vec{A}/(A^j A_j)|$ 
is the distance to the advection control volume center along the direction 
aligned parallel to the cell face normal vector $\vec{A}$ (between neighbor
zone centers).

For advection from one block to another, in order to conserve mass, energy, 
and momentum to round-off instead of truncation, 
it is important {\em not} to interpolate values between ghost zones as was
done for the extrapolated field gradients in the previous section. Instead,
for advection we use the ghost zones as ``buckets'' to capture
material advecting off of the host block. The mass, energy, and momentum
collected in this bucket is then deposited into the corresponding
real zone on the neighboring block that shares a face with the
originating real zone as part of a final loop in the advection routine. 
This is appropriate since zones along inter-block boundaries share faces 
with only a single neighbor.

%\section{Tests}
%\label{sec:tests}

\section{Gradient Test}
\label{sec:gradtest}

Because the cubed-sphere grid uses the same basic gradient operators
that were already tested in Cosmos++ \citep{ann05}, we fully expect
the same second order convergence for smooth fields, at least in the interior
zones. Nevertheless, it is worthwhile to conduct a simple gradient test
for a variety of fields to verify second order convergence over the entire domain,
including at the inter-block boundaries where we have introduced a new
procedure for interpolation of fields beyond local grid domains.

In our first attempt at implementing the cubed-sphere grid, we actually 
did not achieve uniform second order convergence. In that attempt,
instead of constructing the ghost zones as extensions of each block
as described in \S \ref{sec:cubedsphere}, we constructed the ghost
zones to be exact replicas of the nearest zone on the neighboring
block and to mimic the behavior of periodic boundaries on spherical-polar grids. 
However, this introduces a discontinuity into the gradient
operator and actually prevents the convergence of gradients at the
inter-block boundaries. For interior zones {\em not touching an
inter-block boundary}, we found the L1-normalized error for the gradient
of a simple scalar field to converge at second order as expected 
(the L1-normalized error is defined as $E_1 = \sum_{i,j,k} \vert a_{i,j,k} -
A_{i,j,k} \vert/(n_i n_j n_k)$, where $a_{i,j,k}$ and $A_{i,j,k}$
are the numerical and exact solutions, respectively, in each zone
and $n_i$, $n_j$, and $n_k$ are the number of zones in each of the
three directions). However, for
the interior zones {\em touching the inter-block boundaries} (not the ghost
zones themselves, but the zones that touch them), the L1-normalized error {\em did not converge}. 
%($E_1 = [E_1(j=1) + E_1(j=n_j-1) + E_1(k=1) + E_1(k=n_k-1)]/4$,
%where, for instance, $E_1(k=1)$ is the
%L1-normalized error along one of the two-dimensional inter-block faces).
To explain where this failure arises we first note that the 
gradient of a generic field $F$ in
Cosmos++ is calculated as (akin to equation \ref{eqn:fv_adv})
\begin{equation}
G_i = \partial_i F = - \frac{1}{V_z}\sum\limits^\mathrm{faces}_f (F^* A_i)_f ~,
\end{equation}
where the summation is performed over all cell faces. 
%Here $V_z$ is
%the zone volume and $A_j$ is the area vector normal to the cell face
%($f$) pointing inward toward the cell center. 
The problem arises in
calculating $F^*$, the face-centered value of the field; Cosmos++
uses a simple average of the zone-centered values $F_z$ in the two
zones adjoining at face $f$. However, when the line connecting the
two zone centers does not pass through the center of the zone face,
as is the case for nearest neighbor cells across an inter-block
boundary, this simple averaging does not give the correct face-centered 
value $F^*$. In fact, it is relatively easy to show in this
case that the absolute error ($\vert a_{i,j,k} - A_{i,j,k} \vert$) 
in each zone along the inter-block boundary
remains essentially constant,
regardless of the resolution (it only depends weakly on the location
of the zone along the boundary), thus explaining the
non-convergence in these zones. 

The ghost-zone construction described in \S
\ref{sec:cubedsphere}, on the other hand, which is the only one used
for the remainder of this work, restores 2nd order convergence in
all interior zones by giving a properly extrapolated value for $F^*$. Here $F^*=0.5(F_z+F_W)$ is a simple average of the zone-centered value $F_z$ in the interior zone and the ghost-zone weighted average $F_W$ from equation (\ref{eqn:fw}). 
We have confirmed that all interior 
zones (including those touching the inter-block boundaries) give errors at the level of round-off for flat fields and second-order
convergence for all linear and higher-order fields.

%\subsection{Blast-Wave Test}

%JAY?

\section{Tilted Accretion Disks}
\label{sec:disks}

Having demonstrated that our implementation of the cubed-sphere grid 
preserves the correct convergence order for our code, 
%and does not introduce spurious features along the inter-cube boundaries
we can confidently move on to testing our primary application of interest: black-hole accretion disks. We begin with a review of how the simulations are initialized and then consider two sets of test cases: In \S \ref{sec:schw} we study disks of differing alignments relative to a Schwarzschild black hole; in \S \ref{sec:Kerr} we compare tilted disk simulations around a Kerr black hole, one carried out on a spherical-polar mesh and the others on the cubed-sphere.

\begin{figure*}
%\plottwo{Cubed_Sphere_Grid.eps}{Spherical_Polar_Grid.eps} 
\plottwo{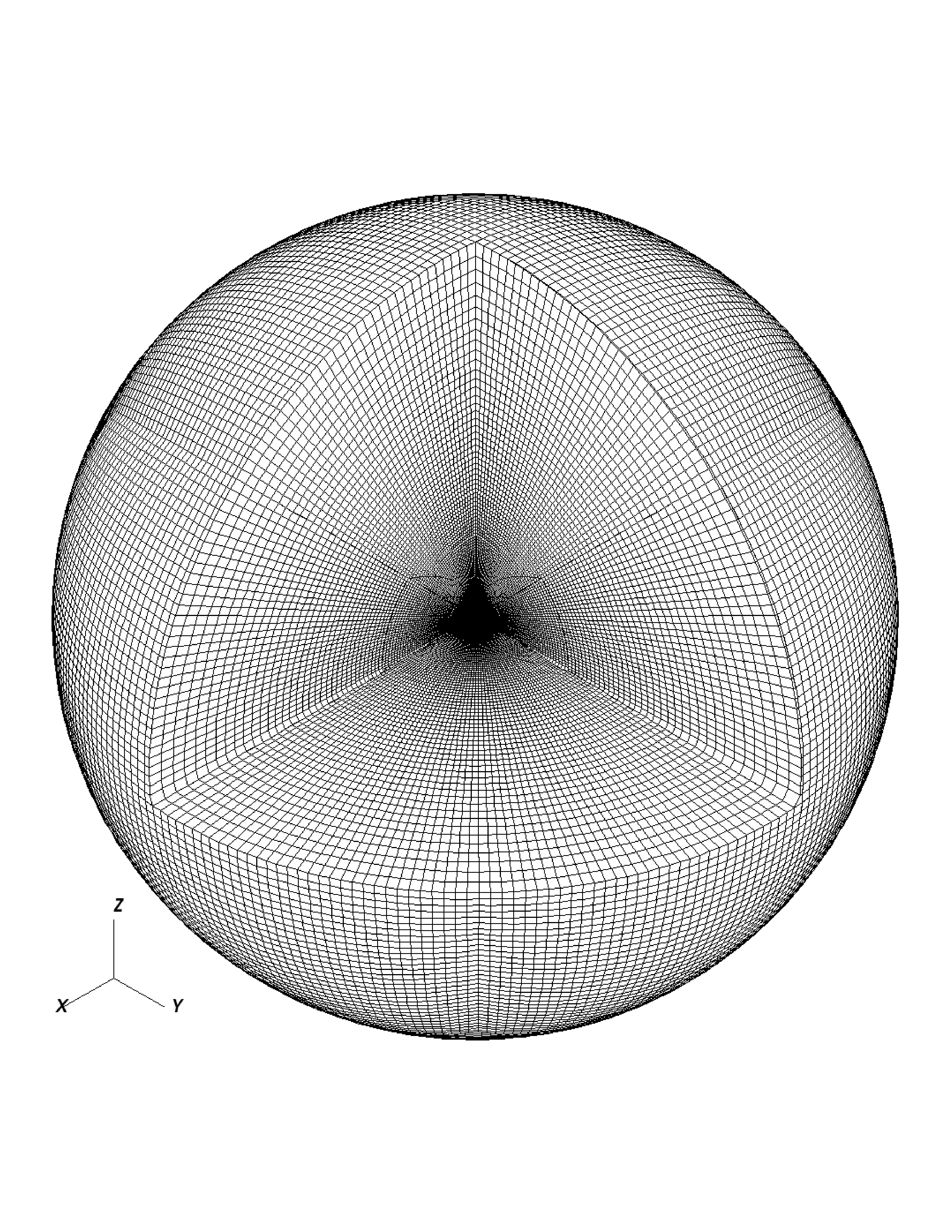}{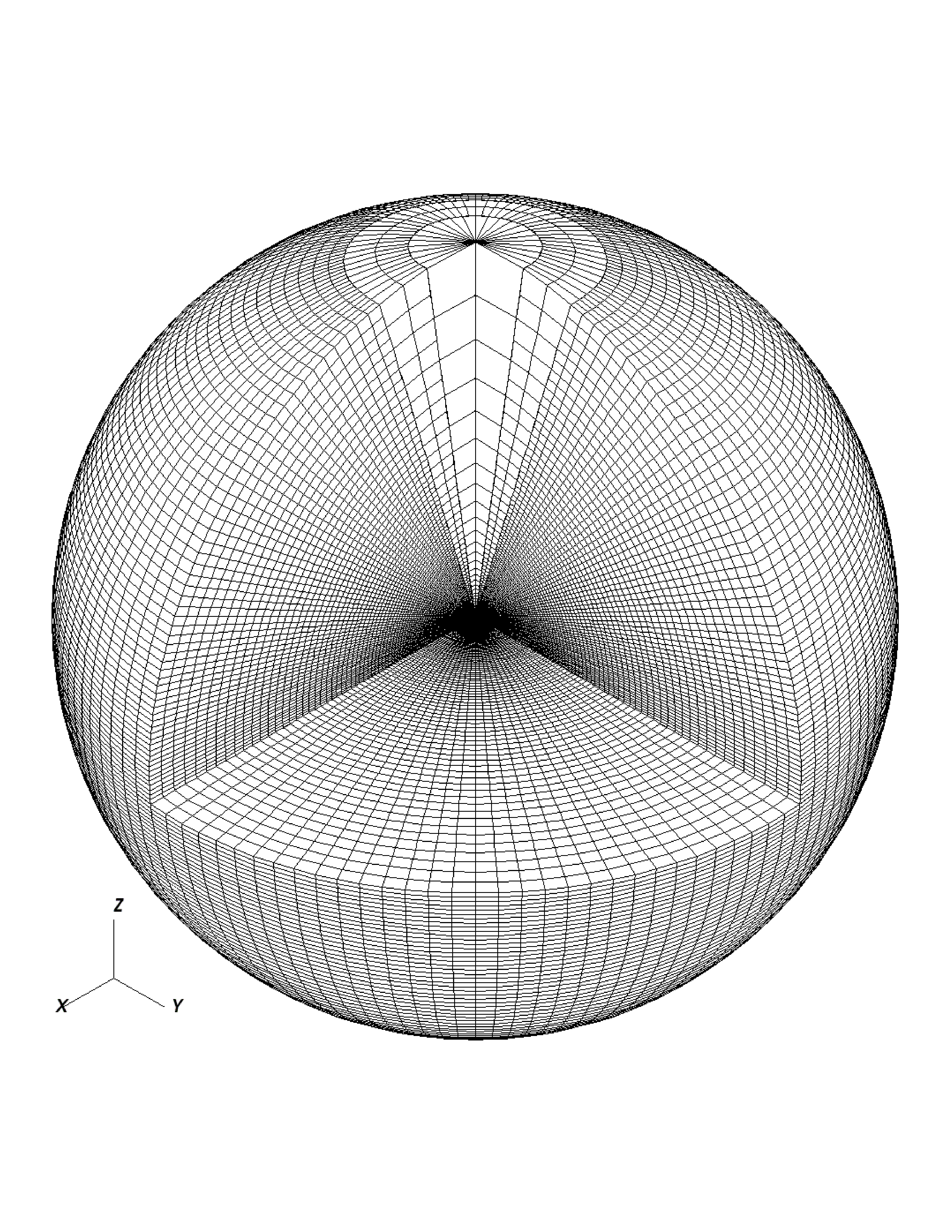} 
\caption{({\em left}) Plot of the cubed-sphere grid geometry used for the
disk simulations presented in this work. ({\em right}) Plot of the spherical-polar grid used in \citet{fra07b}, including an underresolved polar region.
\label{fig:cubedsphere}}
\end{figure*}

Most of the accretion disk simulations presented in this
work using the cubed-sphere grid are carried out at a resolution of
$128\times64\times64\times6$, where there are 128 radial shells and each of the blocks are resolved
with $64\times64$ angular zones. Along its symmetry planes,
such a grid looks like a spherical polar grid of resolution
$128\times128\times256$. However, the more uniform distribution of
zones in the cubed-sphere grid means we are able to achieve such 
resolution with a smaller number of zones overall (by a factor of 3/4). Also, because of the more uniform zone sizing, we are able to run with a Courant time step that is almost 30 times larger than could be used with a spherical-polar grid of that resolution, which means the required CPU cycle-count is smaller by the same factor.
An image of the actual
grid used in these simulations is shown in the {\em left} panel of Figure
\ref{fig:cubedsphere}. This can be compared to the spherical-polar grid used in 
\citet{fra07b}, including the underresolved polar regions, which is shown in the {\em right} panel of Figure \ref{fig:cubedsphere}. The timestep for the cubed-sphere grid is even 25\% larger than for that special grid, where the pole was underresolved precisely to keep the timestep reasonable.

%\clearpage
 
%\clearpage

The inner and outer radial boundaries are set at $0.98 r_\mathrm{BH}$ 
and $120 r_G$, respectively, where $r_\mathrm{BH}$ is the radius of the 
black-hole horizon and $r_G = GM_\mathrm{BH}/c^2$ is the gravitational 
radius. Note that, because we use the Kerr-Schild form of the Kerr metric, we are able to place the inner radial boundary inside the black-hole horizon. In principle, this should keep the inner boundary causally disconnected from the flow, although numerically there is still some communication. At both the inner and outer radial boundaries we apply ``outflow'' conditions: Fluid variables are set the same in the external
boundary zone as in the neighboring internal zone, except for
velocity, which is chosen to satisfy
\begin{equation}
V^r_\mathrm{ext} = \left\{ \begin{array}{cc}
          V^r_\mathrm{int} & V^r \mathrm{~points~off~the~grid}~, \\
         -V^r_\mathrm{int} & V^r \mathrm{~points~onto~the~grid}~.
         \end{array} \right.
\end{equation}

For the initial conditions of the simulations we start from the
commonly used analytic solution for a hydrostatic fluid torus
orbiting the black hole. In this case, we choose the torus
parameters to be the black-hole spin ($a/M_\mathrm{BH}$), the inner radius of
the torus ($r_{\rm in}=15 r_G$), the radius of the pressure maximum
of the torus ($r_{\rm center}=25 r_G$), and the power-law exponent
($q=1.68$) used in defining the specific angular momentum
distribution,
\begin{equation}
\ell = -u_\phi/u_t = k \Lambda^{2-q} ~,
\end{equation}
where $u_\mu = g_{\mu \nu}u^\nu$, $g_{\mu\nu}$ is the 4-metric, and $u^\mu$ is the fluid 4-velocity. We then follow the procedure in \citet{cha85} to solve for the
initial state of the torus. Knowledge of $r_{\rm center}$ leads
directly to a determination of $\ell_{\rm center}$ by setting it
equal to the geodesic value at that radius. The numerical value of
$k$ comes directly from the choice of $q$ and the determination of
$\Lambda_{\rm center}$, where
\begin{equation}
\frac{1}{\Lambda^2} = -\frac{g_{t \phi}+\ell g_{tt}}{\ell g_{\phi
\phi} + \ell^2 g_{t \phi}}~. \label{eqn:Lambda}
\end{equation}
Finally, having chosen $r_{in}$ we can obtain $u_{in}=u_t (r_{in})$,
the surface binding energy of the torus, from $u_t^{-2} =
g^{tt}-2\ell g^{t\phi} +\ell^2 g^{\phi\phi}$.

The solution of the torus variables can now be specified. The
internal energy of the torus is \citep{dev03b}
\begin{equation}
\epsilon(r,\theta) = \frac{1}{\Gamma} \left[ \frac{u_{in}
f(\ell_{in})}{u_t(r,\theta)f(\ell (r,\theta))} \right] ~,
\end{equation}
where $\ell_{in}=\ell(r_{in})$ is the specific angular momentum of
the fluid at the surface and
\begin{equation}
f(\ell) = \left|1-k^{2/n}\ell^\alpha\right|^{1/\alpha} ~,
\end{equation}
where $n=2-q$ and $\alpha=(2n-2)/n$. Assuming an isentropic equation
of state {\em for the initialization only}, the gas pressure and density 
must be related by the expression 
$P=\rho \epsilon(\Gamma-1)=\kappa \rho^\Gamma$, and so the density is 
given by $\rho = \left[
\epsilon(\Gamma-1)/\kappa \right]^{1/(\Gamma-1)}$. We take
$\Gamma=5/3$ and $\kappa=0.01$ (arbitrary units). Finally, the
angular velocity of the fluid is specified by
\begin{equation}
\Omega = V^\phi = -\frac{g_{t\phi}+\ell g_{tt}}{g_{\phi \phi} + \ell
g_{t \phi}} ~.
\end{equation}

The torus is then seeded with weak poloidal magnetic field loops with
non-zero spatial components $\mathcal{B}^r = -
\partial_\vartheta A_\varphi$ and $\mathcal{B}^\vartheta =
\partial_r A_\varphi$, where
\begin{equation}
A_\varphi = \left\{ \begin{array}{ccc}
          b(\rho-\rho_{\rm cut}) & \mathrm{for} & \rho\ge\rho_{\rm cut}~, \\
          0                  & \mathrm{for} & \rho<\rho_{\rm cut}~.
         \end{array} \right.
\label{eq:torusb}
\end{equation}
The parameter $\rho_{\rm cut}=0.5*\rho_{\rm max,0}$ is used to keep
the field a suitable distance inside the surface of the torus initially, where
$\rho_{\rm max,0}$ is the initial density maximum within the torus.
Using the constant $b$ in equation (\ref{eq:torusb}), the field is
normalized such that initially $\beta_{\rm mag} =P/P_B \ge
\beta_{\rm mag,0}=10$ throughout the torus, where $P_B$ is the magnetic pressure. The magnetic field is added in order to seed the magneto-rotational instability
\citep[MRI; ][]{bal91}, which is now commonly believed to be the
source of angular momentum transport within black-hole accretion
disks \citep{bal98}.

As mentioned in \S \ref{sec:cubedsphere}, our implementation of the
cubed-sphere requires that the black-hole be aligned with the grid.
Therefore, unlike our previous work where we tilted the black
hole, if we want a tilted configuration now we must tilt the
disk. By itself, tilting the disk is a rather trivial operation,
simply requiring the following coordinate transformation be applied prior to constructing the torus:
\begin{eqnarray}
x^\prime & = & x\cos \beta_0 - z \sin \beta_0 \nonumber \\
y^\prime & = & y \nonumber \\
z^\prime & = & x\sin \beta_0 + z \cos \beta_0 ~,
\label{eqn:betatransform}
\end{eqnarray}
where $\beta_0$ is the initial tilt of the disk. However, as we
describe in the next section, we were surprised to discover that our
tilted disks evolved differently than our untilted disk, at least at
early times, even for non-rotating, Schwarzschild black holes, for
which a tilt should have no physical meaning or significance.

\subsection{Schwarzschild Black Hole}
\label{sec:schw}

Here we compare simulations of black-hole accretion disks carried out for a Schwarzschild black hole
($a/M_\mathrm{BH}=0$). Table \ref{tab:schw} summarizes the parameters 
for these runs. In our naming convention, the first number indicates the dimensionless spin of the black hole ($a/M$) without the decimal; the second number, if present, gives the tilt angle of the disk in degrees; the final letter is used to distinguish what resolution the simulation is carried out at, ``H'' being our high resolution ($128\times64\times64\times6$) and ``L'' being low ($64\times32\times32\times6$). The two main simulations, 0H and 015H, begin from identical initial
conditions except for the tilt of the disk with respect to the grid, 
which are 0 and $15^\circ$ respectively.  We also include results of a simulation that uses the spherical-polar grid from \citet{fra07b}; this simulation is denoted by the suffix ``SP'' and has an equivalent peak resolution of $128^3$.  We showed in \citet{fra07b} that this was roughly the minimum resolution needed to get a relatively well converged result for this type of problem. Therefore, in the current work, we do not expect our low-resolution simulation (0L) to be converged; they are instead included for the purpose of estimating the rate of convergence when using the cubed-sphere grid.
%The third simulation, 015Hr, 
%begins from the half-way point in
%the evolution of the ``untilted'' simulation (0H), after which we use a 
%remap procedure to tilt everything (by 15 degrees) to match the ``tilted'' 
%simulation (015H). In this remap procedure primitive data are simply read 
%in from simulation 0H, although with their corresponding positions 
%transformed according to equation (\ref{eqn:betatransform}). These data 
%are then redistributed on the new grid at their new positions. The 
%original motivation for doing this was to overcome the discrepancies in 
%the early time evolution between simulations tat started untilted versus 
%ones that started with a tilt (to be described below). An additional 
%potential benefit of the procedure is that multiple tilt angles $\beta$ 
%could be tested using a single remapped data set, without rerunning 
%the first part of the simulation where the tilt has relatively little effect.

\begin{deluxetable*}{ccccccc}
\tabletypesize{\scriptsize}
\tablecaption{Schwarzschild Simulation Parameters \label{tab:schw}}
\tablewidth{0pt}
\tablehead{
\colhead{Simulation} & \colhead{$a/M$} & \colhead{Tilt} &
\colhead{Resolution\tablenotemark{a}} &
\colhead{Start\tablenotemark{b}} & \colhead{End\tablenotemark{b}} & 
\colhead{$\dot{M}$\tablenotemark{c}}\\
\colhead{} & \colhead{} & \colhead{Angle} &
\colhead{} & \colhead{Time} & \colhead{Time} & 
\colhead{}
}
\startdata
0L\tablenotemark{d} & 0 & 0 & $64\times32\times32\times6$ & 0 & 4 & -0.0129 \\
0H\tablenotemark{d} & 0 & 0 & $128\times64\times64\times6$ & 0 & 4 & -0.0090 \\
015H\tablenotemark{d} & 0 & $15^\circ$ & $128\times64\times64\times6$ & 0 & 4 & -0.0085 \\
%015Hr & 0 & $15^\circ$ & $128\times64\times64\times6$ & 2 & 4 \\
0SP\tablenotemark{e} & 0 & 0 & $128^3$ & 0 & 4 & -0.0141 \\
\enddata

\tablenotetext{a}{In the case of the spherical-polar grid this represents the equivalent peak resolution of an unrefined grid.}
\tablenotetext{b}{In units of $t_{\rm orb}=785 GM/c^3$, the geodesic orbital period
at the initial pressure maximum $r_{\rm center}$.}
\tablenotetext{c}{Calculated from the slopes of $M$ vs. $t$ over the interval $3 \le t/t_\mathrm{orb} \le 4$.}
\tablenotetext{d}{Cubed-sphere grid.}
\tablenotetext{e}{Multi-resolution-layer spherical-polar grid.}

\end{deluxetable*}

\begin{figure}
%\plotone{torusCS.d.0x_angMom.eps} 
\plotone{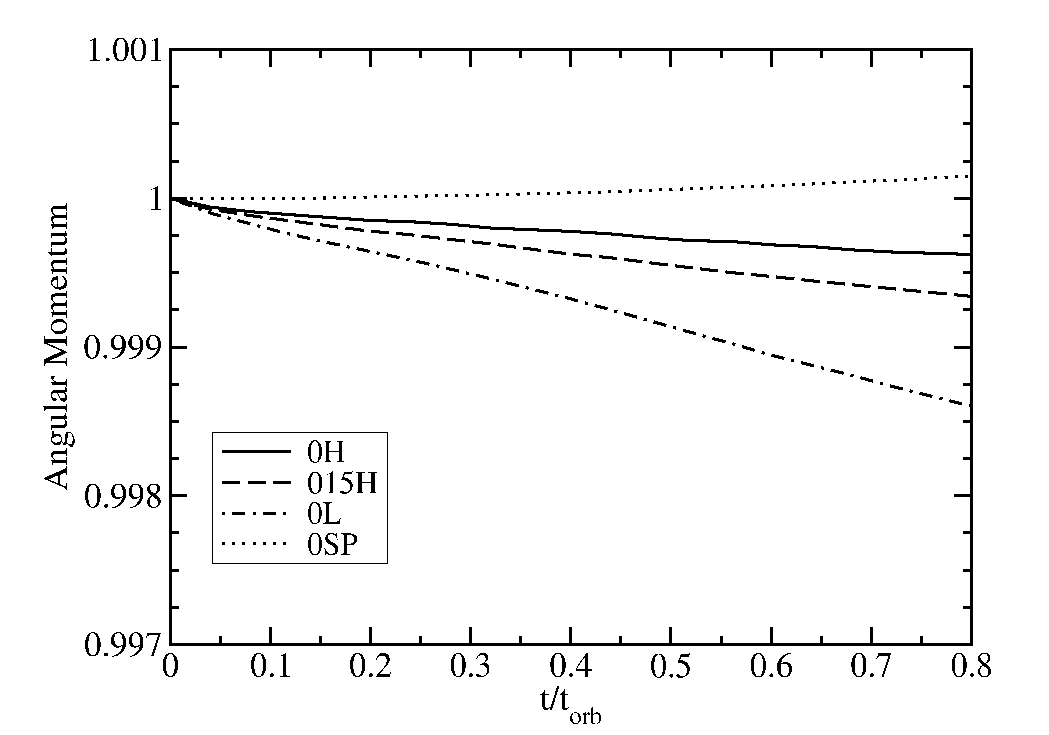} 
\caption{Plot of the total angular momentum as a function of time for simulations 0H ({\em solid}), 015H ({\em dashed}), 0L ({\em dot-dashed}), and 0SP ({\em dotted}). All plots have been normalized to their initial angular momenta. Simulation 0SP uses the spherical-polar grid described in \citet{fra07b}. 
\label{fig:schw_angMom}}
\end{figure}

Our first concern with the cubed-sphere grid is that the angular-momentum 
conservation may not be sufficient for the purpose of following the 
long-term evolution of an accretion disk, particularly as the flow crosses
the coordinate discontinuities at block boundaries. At a minimum we want to 
quantify our angular-momentum-conservation error, which we do graphically 
in Figure \ref{fig:schw_angMom}, where we plot the total angular momentum 
in each simulation as a function of time for runs 0L, 0H, 015H, and 0SP. 
The total angular momentum is defined as 
\begin{equation}
\int_V T^0_{~\phi} \sqrt{-g} \mathrm{d}V ~,
\end{equation}
where $T^0_{~\phi} = (\rho h + 2 P_B)u^0 u_\phi - (B^0 B_\phi)/(4\pi)$, $g$ is the determinant of the 4-metric, $B^\mu$ is the magnetic field 4-vector, and 
\begin{equation}
h = 1 + \epsilon + \frac{P}{\rho}
\end{equation}
is the relativistic enthalpy. We only plot part of the first orbital 
period ($t \le 0.8 t_\mathrm{orb}$) of data because after this time 
significant amounts of angular momentum begin to advect into the 
black hole and leave the outer boundary of the grid in jets and winds, 
so that it is much more difficult to track the global conservation. 
Ideally all the lines in Figure \ref{fig:schw_angMom} would be perfectly horizontal, 
indicating exact angular momentum conservation, but we do not really expect 
this (slight imbalances in the momentum ``source'' terms and imperfect boundary conditions, for instance, can prevent exact conservation). The angular momentum conservation in our ``low'' resolution simulation 
0L is 0.18\% (extrapolated to a full orbital period); this drops down to 
0.048\% per orbital period at our normal resolution, about the level of 
convergence (second order) we expect. Furthermore, it appears that this 
error is not strongly dependent on the orientation of the disk with respect 
to the grid, based on a comparison of simulations 0H and 015H. 
Simulation 0SP is included to give some indication of our typical angular 
momentum conservation error on the multi-resolution-layer spherical-polar 
grid used in our previous work. The error in this case is 0.019\% per orbital period, somewhat better but still comparable to simulation 0H, suggesting we suffer only a small degradation in angular momentum conservation in going from our spherical-polar grid to the cubed-sphere grid.

%\clearpage
 
%\clearpage

\begin{figure*}
%\epsscale{1.}
%\plotone{torusCS.d.0x.Rho_xz_1.eps} 
\plotone{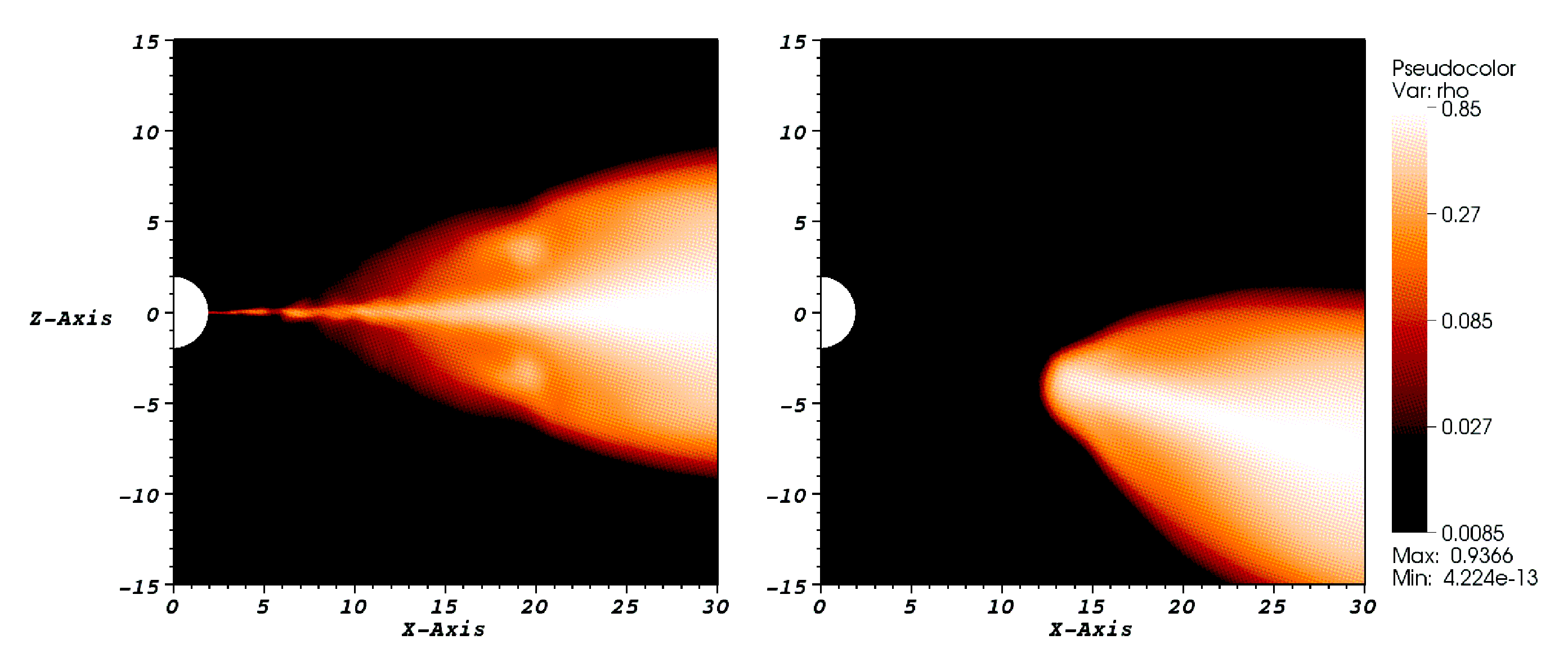} 
\caption{Plot of logarithm of density (normalized to $\rho_{0, \mathrm{max}}$) along an azimuthal slice at $\phi=0$ at $t=1 t_{\rm orb}$ for simulations 0H ({\em left panel}) and 015H ({\em right panel}). Because the black hole is not rotating in this simulation, the tilt should have no physical effect and we would expect the two simulations to evolve nearly identically. The observed differences are due to the numerical treatment of the current sheet that forms in the midplane of the disk, as described in the text.
\label{fig:schw_rho}}
\end{figure*}

Because we are simulating a non-rotating black hole in this section, any tilt we assign the disk has
no physical meaning; it can only be defined relative to the grid. We would expect, therefore, that this tilt would not have any physical effect on the evolution. Interestingly, that is not what we find at early times. The difference is perhaps shown most graphically in Figure \ref{fig:schw_rho}, which shows the gas density of the disk for simulations 0H and 015H along one azimuthal slice after one orbital period ($t_{\rm orb}$) at the initial pressure maximum ($r_{\rm center}$). In simulation 0H ({\em left} panel of Fig. \ref{fig:schw_rho}), the disk has spread radially to such an extent that it reaches all the way to the event horizon of the black hole (inner boundary of the computational grid). In simulation 015H, on the other hand ({\em right} panel), the disk has hardly spread radially at all, having started at $r_{\rm in}=15 r_G$ and only penetrated to $12 r_G$.

The primary mechanism responsible for the radial spreading of the disk over the first orbital period is not solely the MRI, but also the differential winding of the initial radial component of the poloidal field loops, the so-called $\Omega$-dynamo. The amplified toroidal and radial field components allow for efficient angular momentum transport essentially from the beginning of the simulation. This is, of course, peculiar to an initial field configuration such as ours which includes a radial field component. If, instead, we started from a purely toroidal field, differential winding would not play a role initially and angular momentum transport would have to await a more complete development of the MRI, which occurs on roughly an orbital timescale. 

\begin{figure}
%\plotone{torusCS.d.0x_energy.eps} 
\plotone{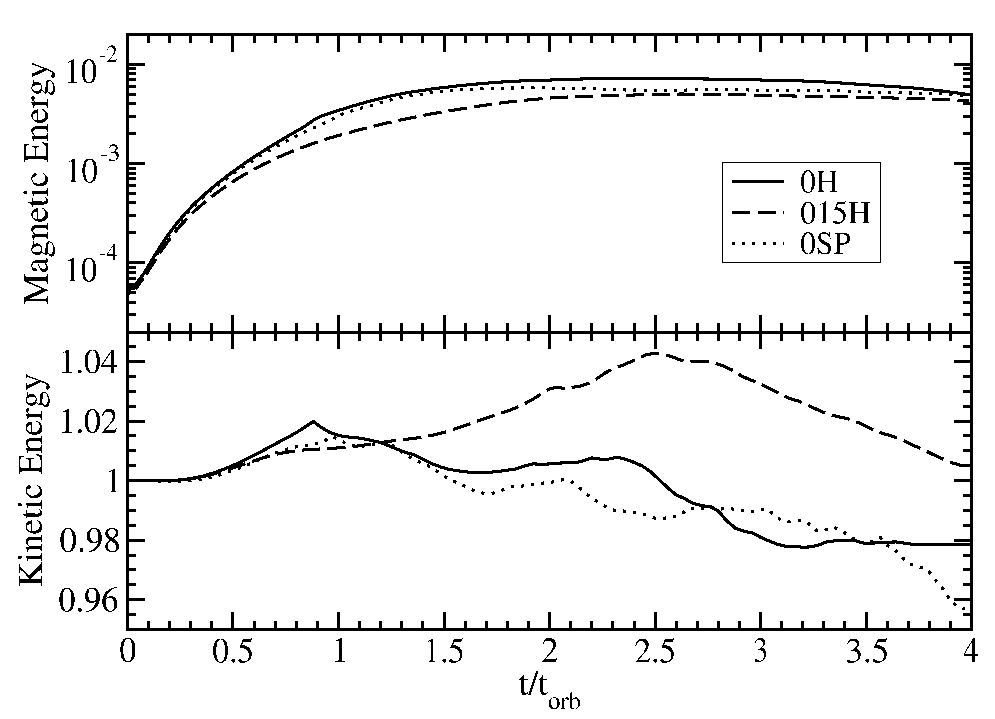} 
\caption{Plot of the total magnetic  ({\em top}) and kinetic ({\em bottom}) energies as functions of time for simulations 0H ({\em solid}), 015H ({\em dashed}), and 0SP ({\em dotted}). All plots have been normalized by the initial kinetic energy of simulation 0H.\label{fig:schw_magE}}
\end{figure}

Something in simulation 015H appears to be shorting out the shear amplification of the field as compared to simulation 0H. Growth of the MRI also appears to be delayed, as evidenced by the less turbulent appearance of simulation 015H in Figure \ref{fig:schw_rho}. This may be related to the lack of an $\Omega$-dynamo since the MRI has less field to grow on whenever this is inactive \citep{haw02}. Furthermore, we can see for certain in Figure \ref{fig:schw_magE} that the total magnetic energy is growing more slowly in simulation 015H than in 0H (and 0SP). Here we define the magnetic and kinetic energies as
\begin{equation}
\sqrt{-g}\left[\left(g^{00}+2 u^0 u^0\right)P_B - \frac{B^0 B^0}{4\pi} \right]
\end{equation}
and $Dh(u^0-1)$, respectively, where $D=W\rho$ is the generalized fluid density with boost $W=\sqrt{-g}u^0$. Both energies are summed over the entire simulation domain. All three simulations show a very rapid initial growth of the magnetic energy due to the combination of shear amplification and the MRI. They also show a gradual increase in kinetic energy over the first orbit as gravitational potential energy is converted into kinetic. After approximately $1t_\mathrm{orb}$ the growth of the magnetic fields saturates. At about the same time in simulations 0H and 0SP, kinetic energy begins accreting into the black hole in significant amounts, accounting for the sudden change in slope. This happens about an orbit and a half later in simulation 015H.

%\clearpage
 
%\clearpage

%[CHRIS: THIS DISCUSSION IS SLIGHTLY OUT OF PLACE/ORDER HERE, MAYBE MENTION IT LATER]
%Figure \ref{fig:schw_magE} also illustrates a couple interesting points about our remapped simulation 015Hr. First, because the remap simply populates zone variables on the new grid with the values corresponding to the nearest zones on the original grid, the procedure does not conserve mass, energy, nor momentum (since zone volumes are likely different on the two grids). This accounts for the discontinuity in kinetic energy between simulations 0H and 015Hr, which is apparent in the {\em bottom} panel of Figure \ref{fig:schw_magE}. It is also apparent from the {\em top} panel that significant magnetic energy is lost after the remap. We also note that the remapped simulation (015Hr) gives a different mass accretion rate at late times (see Table \ref{tab:mdot}). [CHRIS: IS MOST OF THIS LOST OVER THE TORUS, OR DOES IT INCLUDE BACKGROUND GAS?
%THIS COULD LIKELY BE FIXED BY INTERPOLATING CONSERVED QUANTITIES AND DISTRIBUTING
%THEM BY LOCAL VOLUME SCALING - I DID THIS FOR THE COSMOLOGICAL ACCRETION
%PROBLEMS, IT'S SIMPLE ENOUGH TO FIX SO I DON'T THINK WE NEED TO MAKE A BIG
%CASE, MAYBE EVEN DON'T INCLUDE THIS DISCUSSION (OR REMAP RUN/PROCEDURE) AT ALL.]

\begin{figure*}
\epsscale{1.2}
%\plotone{torusCS.d.0x.By_xz_1.eps} 
\plotone{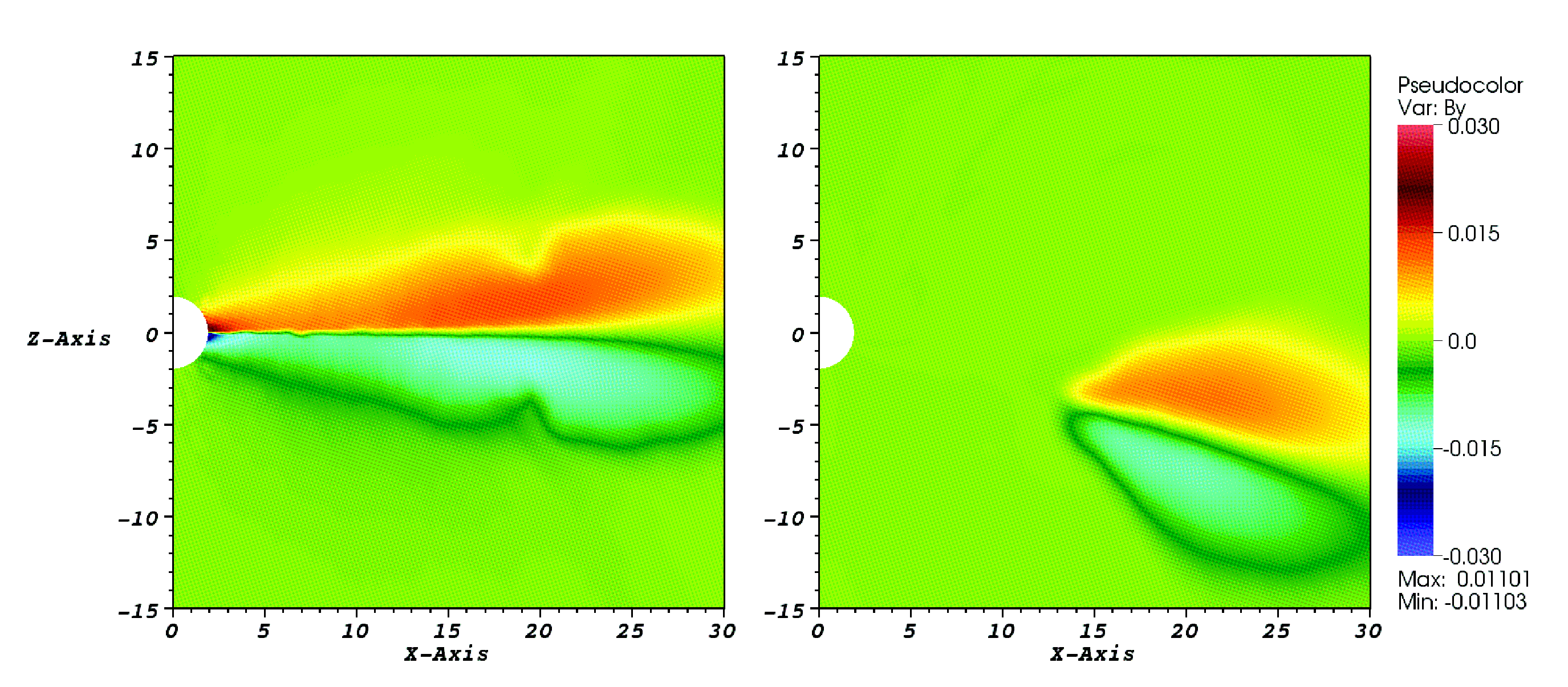} 
\caption{Pseudocolor plot representing the value of $B^y$ (in code units) along an azimuthal slice at $\phi=0$ at $t=1 t_{\rm orb}$ for simulations 0H ({\em left panel}) and 015H ({\em right panel}). The midplane current sheet (represented by the line where the color changes from red to blue) remains essentially sub-zonal in simulation 0H, whereas it is spread across approximately 3 zones in simulation 015H (greenish-yellow zones between red and blue). 
\label{fig:schw_By}}
\end{figure*}

The culprit for the retarded field growth in simulation 015H appears to be the numerical treatment of the current sheet that forms in the midplane of the disk as a result of the differential winding. For an untilted simulation, such as 0H, this current sheet resides almost exactly along an interfacial boundary, right along one of the symmetry planes of the grid (see {\em left} panel of Figure \ref{fig:schw_By}). Furthermore, because this is a nearly perfect symmetry plane for the flow, there is very little advection of fluid across this boundary, and so the current sheet remains relatively stationary. In effect, the current sheet remains unresolved, because it spans less than a full zone's width in the vertical direction. Notice how narrow the current sheet is in the {\em left} panel of Figure \ref{fig:schw_By}. This is not the case for a tilted-disk simulation. By necessity, the disk midplane in a tilted disk is no longer aligned with any symmetry plane of the grid ({\em right} panel of Figure \ref{fig:schw_By}). This means that the disk midplane, and more importantly the midplane current sheet, passes through the interiors of some zones rather than always along their boundary. Numerically this is a critical distinction. For a zone-centered code such as Cosmos++ whenever a current sheet is aligned along an interfacial boundary that experiences no advection, as is approximately the case in our untilted simulations (0H and 0SP), there can be no numerical reconnection and magnetic fields are preserved. If, on the other hand, the current sheet passes through a zone center, as it does in our tilted simulation (015H), numerical reconnection is greatly enhanced. The effect is to drain energy from the magnetic field. In the present work, which uses the internal energy conserving form of Cosmos++ this energy is simply lost from the simulations (see \citet{fra08a} for a discussion of the implications of the different forms of energy conservation in numerical simulations of black-hole accretion disks).

%\clearpage
 
%\clearpage

\begin{figure*}
\epsscale{1.2}
%\plotone{torusCS.d.0x.Rho_xz_4.eps} 
\plotone{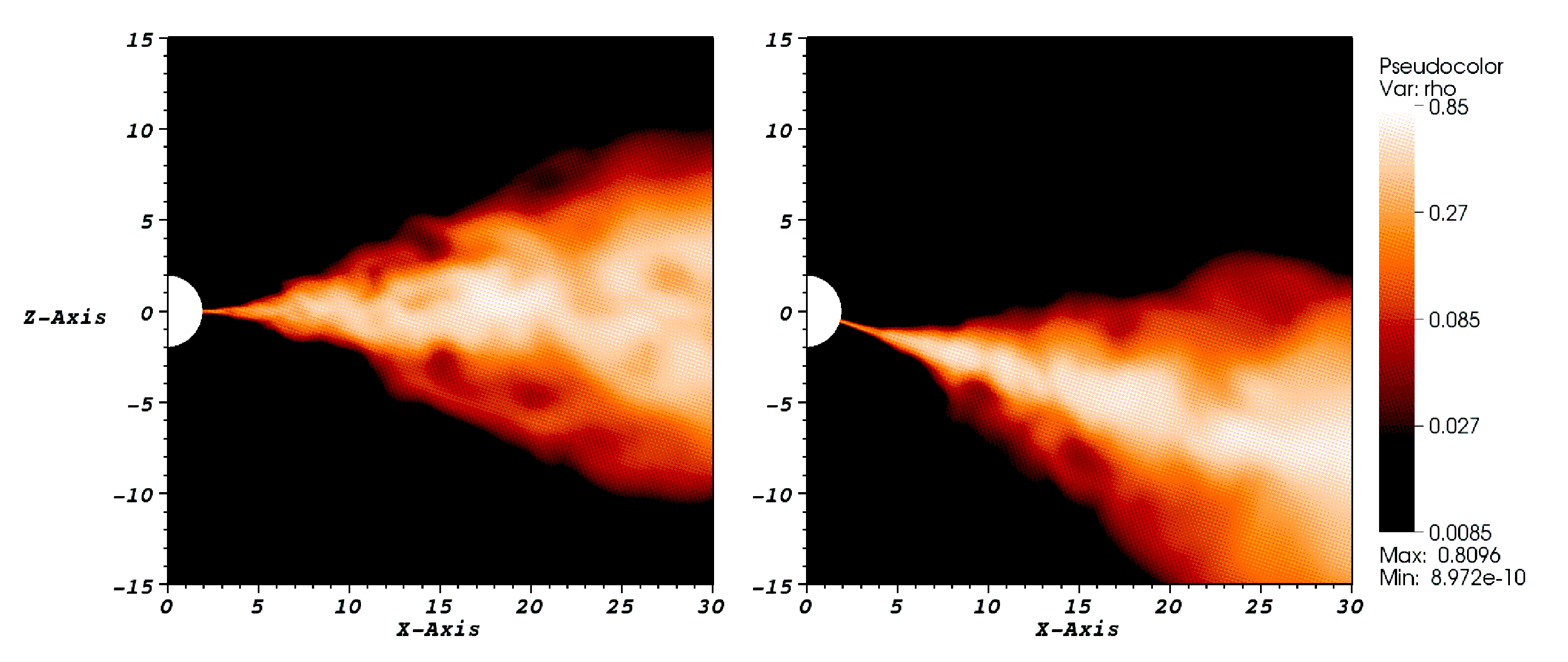} 
\caption{Same as Fig. \ref{fig:schw_rho}, except at time $t=4t_\mathrm{orb}$ instead of $1t_\mathrm{orb}$. Here the discrepancies between the untilted ({\em left panel}) and tilted ({\em right panel}) simulations are greatly diminished. \label{fig:schw_rho2}}
\end{figure*}

This is a somewhat worrisome discovery; however, we emphasize that it is restricted to the particular field geometry we start with, as no strong midplane current sheet forms if one starts from a purely toroidal field. Furthermore, as the disk becomes more turbulent with the action of the MRI, we find that the discrepancies between the tilted and untilted simulations are dramatically reduced to the point that, at late times, they are nearly indistinguishable. For instance, in Figure \ref{fig:schw_rho2}, we show plots equivalent to Figure \ref{fig:schw_rho}, except at $t=4t_\mathrm{orb}$ as opposed to $1t_\mathrm{orb}$, which show the two disks to be nearly identical. The late-time mass accretion rates are also quite similar (see Table \ref{tab:schw}).

%\clearpage
%\clearpage

\begin{figure*}
\epsscale{1.}
%\plotone{torusCS.d.0x_diskProperties.eps}
\plotone{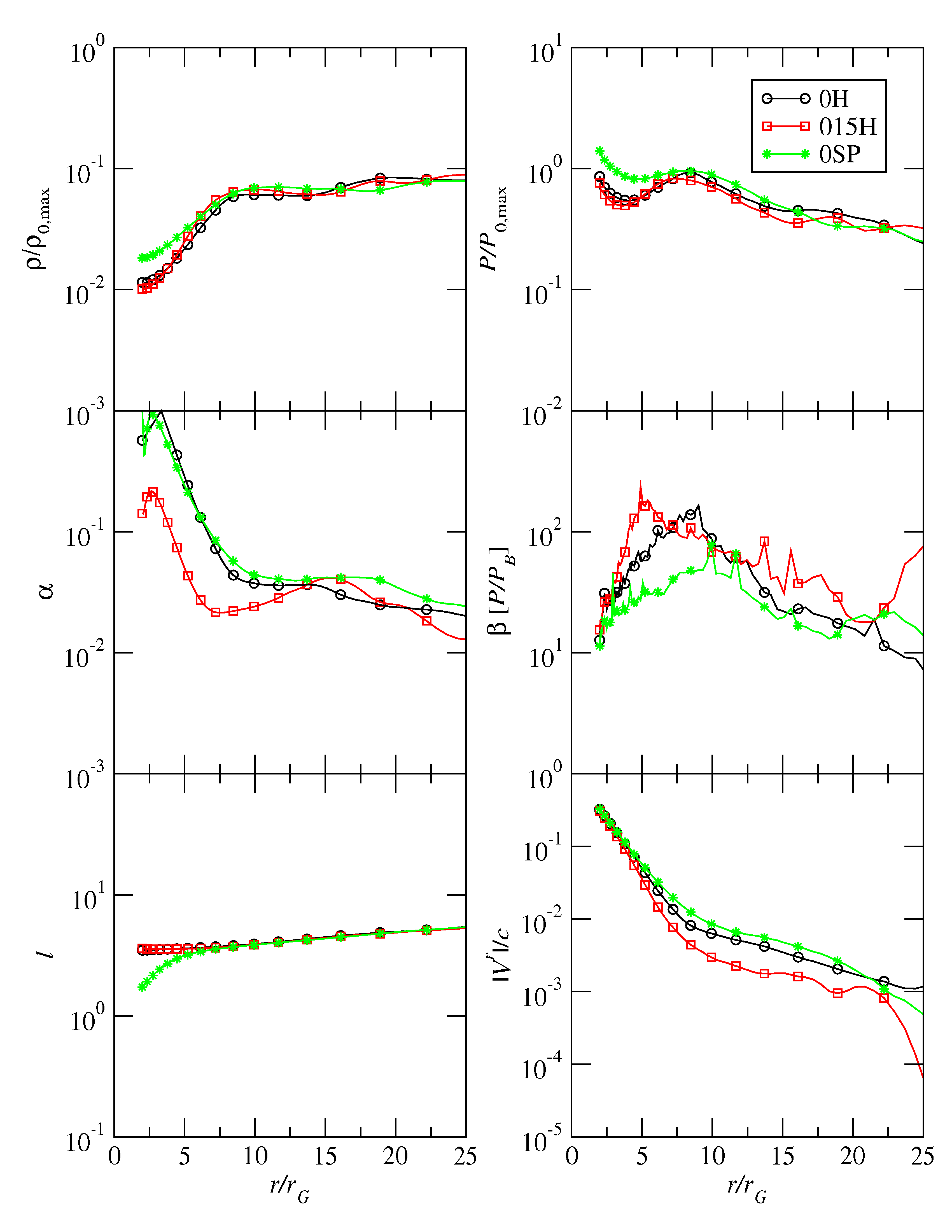}
\caption{Main disk properties plotted as a function of radius for
simulations 0H, 015H, and 0SP. The data have been time-averaged over
the final orbital period of each simulation ($3 \le t/t_\mathrm{orb} \le 4$). $P$, $\alpha$, $\beta$, $\ell$, and $V^r$ are mass-weighted averages of the pressure, dimensionless stress, plasma equipartition parameter, specific angular momentum, and radial inflow velocity, respectively.
\label{fig:schw_properties}}
\end{figure*}

For a more rigorous comparison, in Figure \ref{fig:schw_properties}, we present time- and shell-averaged values of density ($\rho$), gas pressure ($P$), dimensionless stress ($\alpha$), plasma magnetization parameter ($\beta_\mathrm{mag}$), specific angular momentum ($\ell$), and radial inflow velocity ($\overline{V}^r$) as functions of radius for simulations 0H, 015H, and 0SP at late time, where
\begin{equation}
\alpha = \left\langle \frac{ \vert u^r u^\varphi \vert\vert B
\vert\vert^2 - B^r B^\varphi \vert}{4 \pi P} \right\rangle 
\end{equation}
and $\overline{V}^r=\langle \rho V^r
\rangle/\langle \rho \rangle$. Angle brackets indicate that a radial shell-average has been taken, where 
\begin{equation}
\langle\mathcal{Q}\rangle(r,t) = \frac{1}{A} \int^{2\pi}_0
\int^\pi_0 \mathcal{Q} \sqrt{-g}
\, \mathrm{d}\theta \, \mathrm{d}\phi ~,
\end{equation}
and $A = \int^{2\pi}_0 \int^\pi_0 \sqrt{-g} \, \mathrm{d}\theta
\, \mathrm{d}\phi$ is the surface area of a given radial shell. The time-averaging is done over the interval $3 \le t/t_\mathrm{orb} \le 4$. The shell-averages for $P$, $\alpha$, $\beta_\mathrm{mag}$, $\ell$, and $\overline{V}^r$ are mass-weighted. Measurements of $\rho$, $P$, and $\ell$ show very good agreement between all three simulations, with errors everywhere $\lesssim 20$\% and generally much less. The discrepancies in $\alpha$, $\beta_\mathrm{mag}$, and $\overline{V}^r$ are similarly small for simulations 0H and 0SP, but considerably larger for simulation 015H. This is not unexpected as these quantities depend sensitively on the distribution of magnetic field, meaning they are more affected by the delayed growth of the MRI. 
%Finally, simulations 0H and 015H show very good agreement in $\ell$, but deviate considerably from simulation 0SP, which shows a dramatic drop in the value of $\ell$ inside the marginally stable orbit radius ($r_{ms}=6 r_G$). A similar difference is seen in comparing $\ell$ for the Kerr simulations below. This difference is potentially important as the behavior of $\ell$ inside the marginally stable orbit radius bears on discussions of where the true inner edge of the accretion disk lies \citep[see e.g.][]{kro02} and how much energy can be radiated from the disk before it plunges into the black hole. Truly understanding the difference in behavior of $\ell$ will require careful analysis and comparison that goes beyond the intent of this paper. We, therefore, leave it for future work.

%\clearpage
 
%\clearpage

%For post-processing the tilted data we have to transform the disk
%back into the symmetry plane of the black hole so that variable such
%as $\alpha$ and $\ell$ have the same meaning in both the tilted and
%untilted simulations.

\subsection{Kerr Black Hole}
\label{sec:Kerr}

Having shown that the late-time evolution of simulated black-hole 
accretion disks on our cubed-sphere grid is relatively independent 
of the orientation of the disk with respect to the grid by analyzing 
a few Schwarzschild test cases, we can now evaluate the treatment of 
tilted accretion disks around modestly rotating ($a/M_\mathrm{BH}=0.5$) 
Kerr black holes. Here our test simulations (515L and 515H), which use the new 
cubed-sphere grid, are compared to a reference simulation (515SP), which 
uses the multi-resolution-layer spherical-polar grid from \citet{fra07b} (shown in Figure \ref{fig:cubedsphere} {\em right} panel). All simulations have an initial tilt angle $\beta_0=15^\circ$.  Again we do not expect the low-resolution simulation (515L) to be converged; instead it is included to provide some indication of the rate of convergence.
%For the cubed-sphere simulation (515H), we use the same remap procedure described in the previous section, wherein we evolve an untilted disk for $2t_\mathrm{orb}$ and then tilt the entire simulation with respect to the grid (and black hole). The simulation is then evolved for an additional $8t_\mathrm{orb}$. 
The parameters for each run are described in Table \ref{tab:Kerr}.

\begin{deluxetable}{cccccc}
\tabletypesize{\scriptsize}
\tablecaption{Kerr Simulation Parameters \label{tab:Kerr}}
\tablewidth{0pt}
\tablehead{
\colhead{Simulation} & \colhead{$a/M$} & \colhead{Tilt} &
\colhead{Resolution\tablenotemark{a}} &
\colhead{End\tablenotemark{b}} & \colhead{$\dot{M}$\tablenotemark{c}}\\
\colhead{} & \colhead{} & \colhead{Angle} &
\colhead{} & \colhead{Time} & \colhead{}
}
\startdata
515L\tablenotemark{d} & 0.5 & $15^\circ$ & $64\times32\times32\times6$ & 10 & -0.0032 \\
515H\tablenotemark{d} & 0.5 & $15^\circ$ & $128\times64\times64\times6$ & 10 & -0.0114 \\
515SP\tablenotemark{e} & 0.5 & $15^\circ$ & $128^3$ & 10 & -0.0122 \\
\enddata

\tablenotetext{a}{In the case of the spherical-polar grid this represents the equivalent peak resolution of an unrefined grid.}
\tablenotetext{b}{In units of $t_{\rm orb}=789 GM/c^3$, the geodesic orbital period
at the initial pressure maximum $r_{\rm center}$.}
\tablenotetext{c}{Calculated from the slopes of $M$ vs. $t$ over the interval $3 \le t/t_\mathrm{orb} \le 4$.}
\tablenotetext{d}{Cubed-sphere grid.}
\tablenotetext{e}{Multi-resolution-layer spherical-polar grid.}

\end{deluxetable}

\begin{figure*}
%\plotone{torus.5x_diskProperties.eps}
\plotone{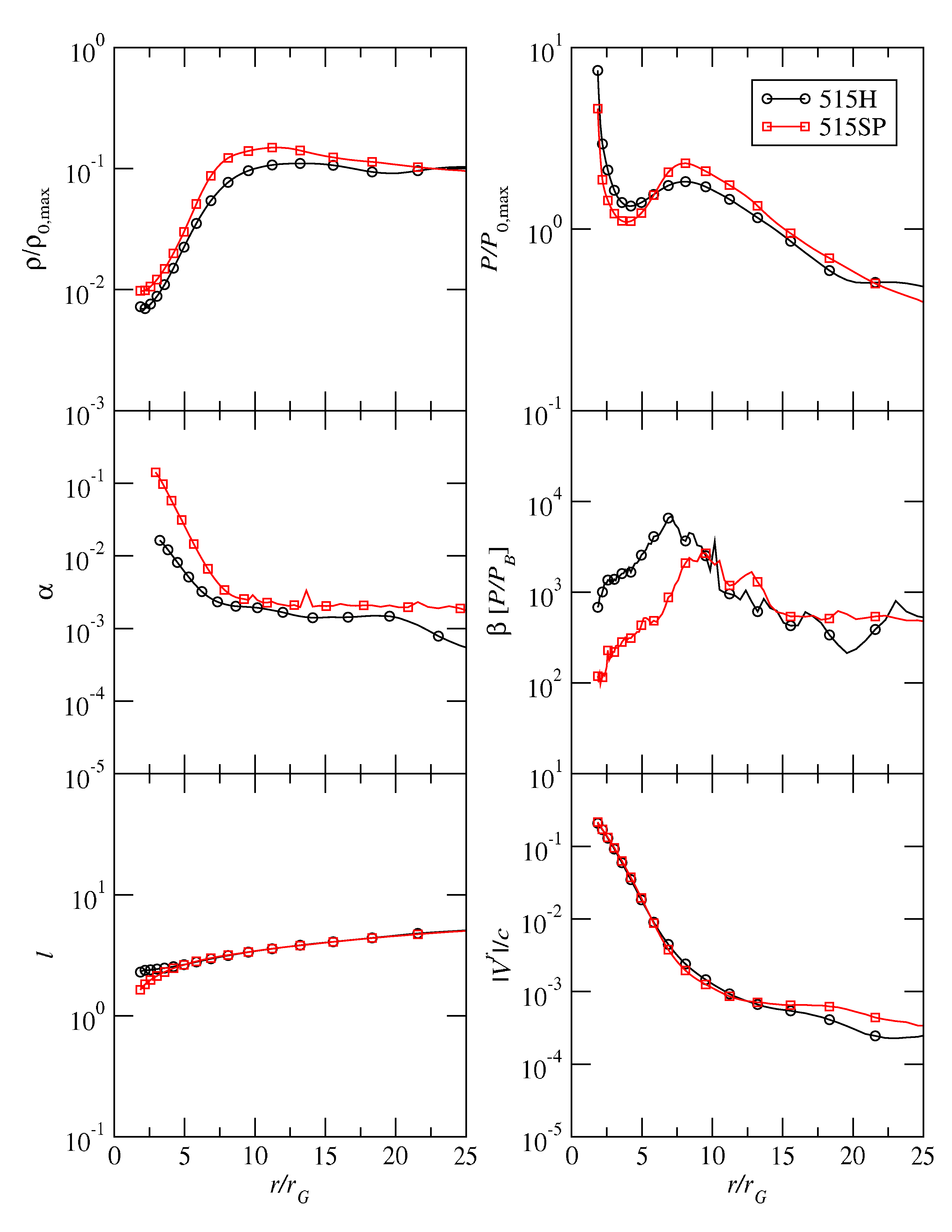}
\epsscale{1}
\caption{Main disk properties plotted as a function of radius for
simulations 515H and 515SP. The data have been time-averaged over
the final two orbital periods of each simulation ($8 \le t/t_\mathrm{orb} \le 10$). $P$, $\alpha$, $\beta$, $\ell$, and $V^r$ are mass-weighted averages of the pressure, dimensionless stress, plasma equipartition parameter, specific angular momentum, and radial inflow velocity, respectively.
\label{fig:Kerr_properties}}
\end{figure*}

First we show in Figure \ref{fig:Kerr_properties} that the general disk properties of simulations 515H and 515SP are quite similar. Again, the largest discrepancies are in the dimensionless stress $\alpha$ and the plasma magnetization parameter $\beta_\mathrm{mag}=P/P_B$. This is not surprising since both of these properties have been shown in previous studies to be quite sensitive to resolution \citep{haw96,fro07}, and although the total number of zones in these two simulations is comparable, the distribution of those zones is considerably different. The level of agreement in the other parameters is really quite remarkable given the very different structures of the grids. Of course, this was exactly what we were hoping to see.
%\clearpage
 
%\clearpage

\begin{figure}
%\plotone{BetavsR_av.eps}
\plotone{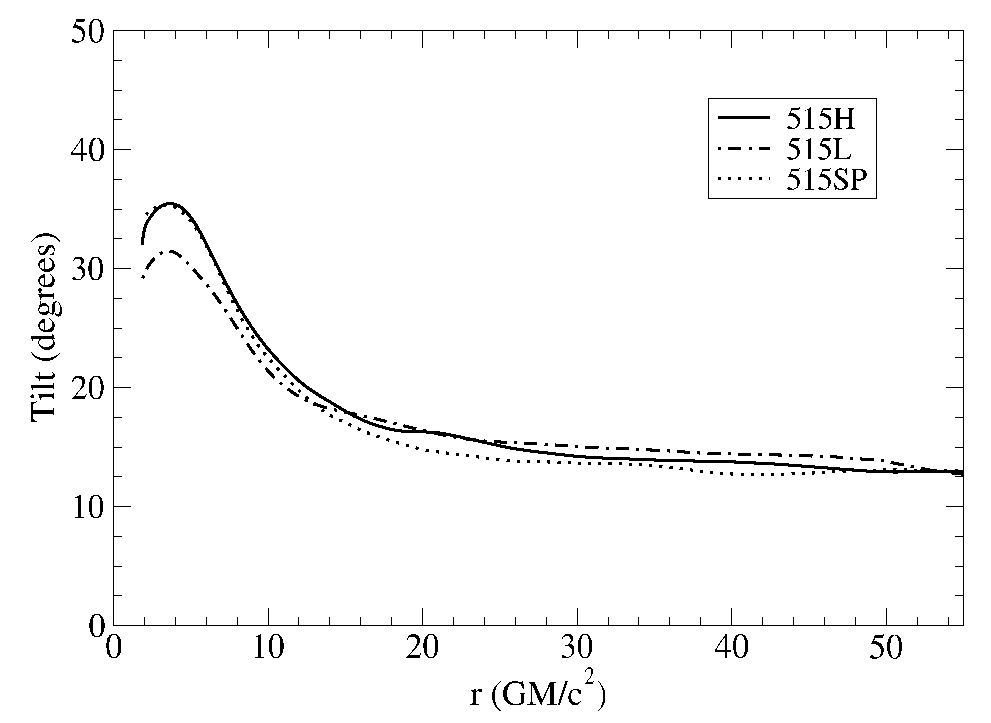}
\caption{Plot of the tilt $\beta$
as a function of radius through the disk for simulations 515L, 515H, and 515SP. 
The data for this plot has been time averaged
over the final two orbital periods of each simulation ($8 \le t/t_\mathrm{orb} \le 10$). The initial tilt
was $\beta_0=15^\circ$. 
\label{fig:beta}}
\end{figure}

Now, because the black-hole is rotating, the tilt of the disk has some physical meaning and consequently causes changes in its evolution relative to an untilted disk, as described in \citet{fra07b} and \citet{fra08b}. For instance, although the disk begins with a uniform tilt of $\beta_0=15^\circ$, we expect a warp caused by the gravitomagnetic torque of the black hole to propagate through the disk as a bending wave. This will cause the tilt $\beta$ to become an oscillating function of radius \citep{iva97,lub02}. In Figure \ref{fig:beta}, we plot $\beta(r)$, time
averaged over the interval $8 t_{\rm orb} \le t \le 10 t_{\rm orb}$, for simulations 515L, 515H, and 515SP.
As in previous work \citep{fra05,fra07b}, we recover the tilt using the definition
\begin{equation}
\beta(r) = \arccos\left[ \frac{\mathbf{J}_{\rm BH} \cdot
\mathbf{J}_{\rm Disk}(r)} {\vert\mathbf{J}_{\rm BH} \vert
\vert\mathbf{J}_{\rm Disk}(r) \vert} \right] ~,
\end{equation}
where $\mathbf{J}_{\rm BH}$ is the angular momentum vector of the black hole and
$\mathbf{J}_{\rm Disk}(r)$ is the angular momentum vector of each radial shell of the simulation domain (dominated by the disk).
Again simulations 515H and 515SP produce remarkably similar results, with discrepancies no larger than $\sim10$\% and generally much smaller. The discrepancies likely have their root in the small differences in conditions at the inner edge of the disk (see Figure \ref{fig:Kerr_properties}) where the bending waves are launched. The 515L simulation exhibits considerably larger discrepancies over most of the disk as expected.

%\clearpage
 
%\clearpage

\begin{figure}
%\plotone{gamma_vs_t.eps}
\plotone{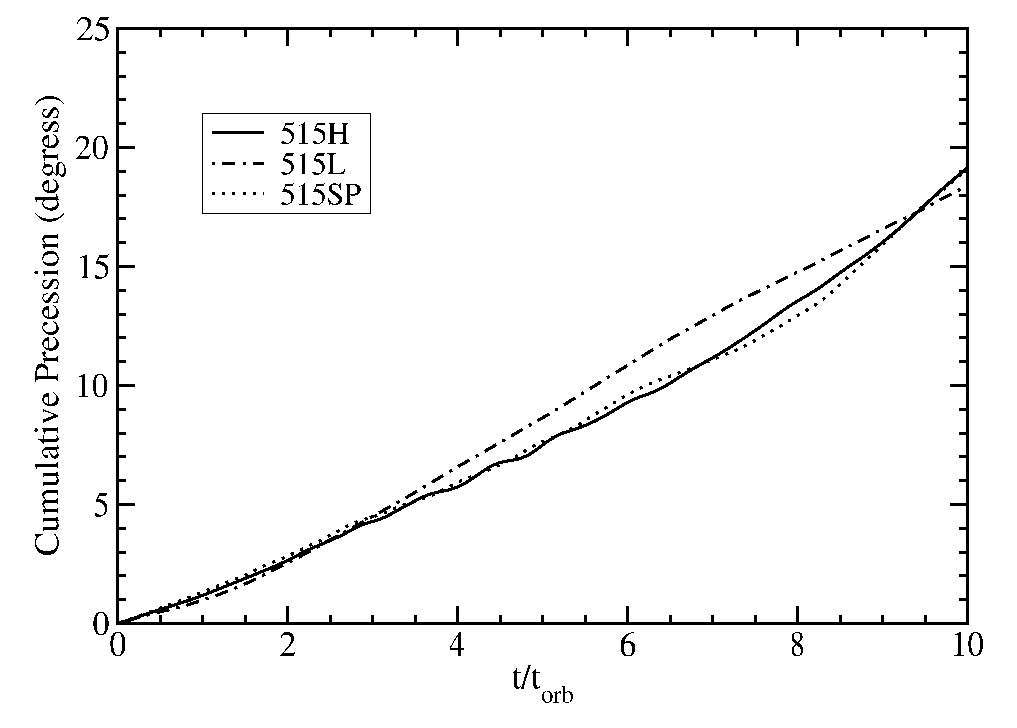}
\caption{Plot of the precession (twist) $\gamma$ as a function of time for simulations 515L, 515H, and 515SP. The slope of this plot can be used to estimate the precession period
of the disk as a whole, which is $0.7 (M/M_\odot)$ s.
\label{fig:precession}}
\end{figure}

Along with warping the disk, the gravitomagnetic torque of the black hole also causes it to precess, particularly in disks such as the ones in our simulations where the fast sound-crossing time causes the disk material to be tightly coupled and respond globally to the torque of the black hole. Global precession of this nature has been noted before in low Mach
number hydrodynamic \citep{fra05} and MHD \citep{fra07b} disks. We track the overall precession
(twist), defined as
\begin{equation}
\gamma = \arccos\left[ \frac{\mathbf{J}_{\rm BH} \times
\mathbf{J}_{\rm Disk}} {\vert \mathbf{J}_{\rm BH} \times
\mathbf{J}_{\rm Disk} \vert} \cdot \hat{y}\right] ~, \label{eq:twist}
\end{equation}
where $\mathbf{J}_{\rm Disk}$ is the total angular momentum vector of the disk and $\hat{y}$ is the unit vector that points along the initial line-of-nodes between the black-hole symmetry plane and disk midplane. In order to capture twists larger than $180^\circ$, we also track
the projection of $\mathbf{J}_{\rm BH} \times \mathbf{J}_{\rm Disk}$ onto
$\hat{x}$, allowing us to break the degeneracy in $\arccos$. By plotting the cumulative precession as a function of time as we have done in Figure \ref{fig:precession}, we make it easy to calculate the precession period of the disk -- in this case $0.7 (M/M_\odot)$~s, which agrees nicely with our predictions for a black-hole of this spin \citep{fra07b}. 

%\clearpage
 
%\clearpage

\section{Conclusion}

In this paper we have presented our implementation of the cubed-sphere grid within Cosmos++. The cubed-sphere grid has at least three significant advantages over more-traditional grid options: 1) it has topological properties similar to a Cartesian grid, but generally conserves angular momentum much better (and nearly as well as a spherical-polar mesh); 2) it can run at a larger Courant-limited timestep than a spherical-polar mesh at comparable resolution (almost a factor of 30 at the resolution used in this work); and 3) it distributes zones more evenly than a spherical-polar mesh, which is desirable for problems where the symmetry is imperfect, such as in tilted accretion disks around rotating black holes, a problem of particular interest to us.
%Our motivation for adding this grid constructor 
%to our code is to study the evolution of jets from tilted black-hole accretion 
%disks, which cannot be done as easily at comparable resolution with the
%more traditional spherical-polar mesh. Although we leave the specific treatment of jets for a future paper, in this work we have clearly demonstrated that the cubed-sphere grid is a viable option for tilted accretion disks around rotating black holes.

In Section \ref{sec:cubedsphere} and Appendix \ref{app:transform} we gave detailed prescriptions for the construction of the cubed-sphere grid and shared a few ``lessons learned'' in regards to extrapolating fields at inter-block boundaries and applying limiters to the field gradients in advection. After implementing these lessons ourselves, we found we could recover second order convergence over the entire grid, including along the inter-block boundaries.

To specifically demonstrate that the cubed-sphere grid is a viable option for the black-hole accretion disk work we have in mind, we have carried out a series of such simulations on our new cubed-sphere grid, using results from our spherical-polar grid as a reference standard. From these tests we conclude that: 

\begin{itemize}
\item{The cubed-sphere grid conserves angular momentum nearly as well a spherical-polar grid at comparable resolution.} 
\item{The angular momentum conservation error on the cubed-sphere grid is only weakly dependent on the tilt of the disk.}
\item{Results on the cubed-sphere grid converge to the same solutions obtained on a spherical-polar grid when the two grids approach comparable resolutions. This is true for both untilted {\em and} tilted disks.}
\item{Important disk properties such as density, pressure, specific angular momentum, inflow velocity, tilt and twist agree to better than 10-20\% for simulations carried out on cubed-sphere and spherical-polar grids with roughly (2-3)$\times10^6$ zones.}
\end{itemize}

During our testing, we made one surprise discovery -- that the early-time evolution was considerably different between our untilted and tilted simulations on the cubed-sphere grid. We found this to be true even for non-rotating Schwarzschild black holes, for which a tilt should have no physical meaning or significance. This is something we had not seen on the spherical-polar grid, but there we had tilted the black-hole, not the disk as we do now. We did not anticipate how important this difference would be for the initial growth and development of the $\Omega$-dynamo and MRI.

We attribute the disparate early-time behavior to the differing ways in which the strong initial current
sheet in our disk is handled numerically when it is tilted with respect to the grid. This is another reminder of the important role that numerical reconnection plays in the evolution of numerically simulated magnetized flows even though this topic is perhaps not given enough emphasis in the literature. The appearance of current sheets is virtually unavoidable in strongly sheared MHD flows such as accretion disks. One possible technique for treating the current sheets more consistently throughout the simulation may be to use an artificial resistivity. This would ensure that the current sheets are always resolved in a similar fashion regardless of their orientation with respect to the grid. However, this technique has only been implemented very recently in relativistic MHD \citep{kom07}. An alternative, although only partial, solution might be to use a total-energy conserving scheme instead of the internal-energy conserving one used here. This, at least, guarantees that the energetics of the flow remain consistent by recapturing in the form of thermal energy any energy lost through magnetic reconnection. When coupled with a radiative cooling package, this can give a much more 
physical description of the evolution of the flow \citep{fra08a}.

%Another way to avoid the problems demonstrated in \S \ref{sec:schw} would 
%be to prevent the formation of a strong midplane current sheet in the 
%first place. This could be accomplished, for instance, by starting with 
%a purely toroidal magnetic field. It has been shown, however, that such 
%a field configuration does not lead to the formation of jetted outflows 
%\citep{bec08}, and since our purpose in implementing the cubed-sphere grid 
%is to study jets from tilted accretion disks, switching to a purely 
%toroidal initial field configuration would be self-defeating.

%Our tests of the Schwarzschild black-hole disks in \S \ref{sec:schw} also pointed out flaws with our remap procedure. Although such a procedure offers advantages in speeding up simulations that have the same black-hole spin but different disk tilts, as implemented it appears to dramatically alter the evolution of the disk. Until a better procedure can be implemented, we will continue starting each new simulation from scratch.

Although the numerical treatments of current sheets and reconnection are important to understand and appreciate, it is equally important in the context of this paper to point out that we demonstrated by numerical example that the long-term evolution of our disks is relatively unaffected by whether or not they are tilted with respect to the grid. As expected, only when the tilt is relative to a rotating black hole are there long-term implications within the disk. 

We are not surprised to find significant discrepancies between our ``low'' and ``high'' resolution simulations, as previous experience had shown us that $128^3$ was roughly the minimum resolution necessary to follow the evolution of black-hole accretion disks in global general-relativistic MHD simulations such as these. Below that resolution the characteristic MRI wavelength ($\lambda_\mathrm{MRI} \equiv 2\pi v_A/\Omega$, where $v_A$ is the Alfv\'en speed) is not covered by a sufficient number of zones over much of the disk volume. This has nothing to do with the cubed-sphere grid itself, but is rather a universal constraint for these types of problems.

Overall we consider our experimentation with the cubed-sphere grid to be a success. In future work we will present further analysis of tilted disks (and their associated jets) evolved using this new grid option.

\begin{acknowledgements}
We would like to recognize Joseph Niehaus for his help testing the cubed-sphere grid. 
We gratefully acknowledge the support of Faculty Research
and Development grants from the College of Charleston, SURF and RPG
grants from the College of Charleston 4th Century Initiative
Program, and a REAP grant from the South Carolina Space Grant
Consortium. A portion of this work was performed under the auspices of the U.S. Department of
Energy by Lawrence Livermore National Laboratory under Contract
DE-AC52-07NA27344. 
%This material is based upon work supported by the National Science 
%Foundation under the following NSF programs: Partnerships for Advanced 
%Computational Infrastructure, Distributed Terascale Facility (DTF) and 
%Terascale Extensions: Enhancements to the Extensible Terascale Facility. 
This work was supported by the National Science Foundation through TeraGrid resources provided by the Texas Advanced Computing Center (TACC). This work also made use of computing resources provided by
the Barcelona Supercomputing Center under activity AECT-2007-3-0002.
\end{acknowledgements}

\appendix
\section{Cubed-Sphere Transformations}
\label{app:transform}

Included in this appendix are the transformations necessary to go from the cubed-sphere coordinates $\{r,\phi_1,\phi_2\}$ to the corresponding spherical-polar ones $\{r,\theta,\phi\}$ on each block.
\begin{itemize}
\item{Block 0 -- (centered about the $+x$-axis; $\pi/4 \le \phi_1 \le 3\pi/4$; $-\pi/4 \le \phi_2 \le \pi/4$)}
\begin{eqnarray}
\cos \theta & = & \frac{\cos \phi_1 \cos \phi_2}{\sqrt{1-(\cos \phi_1 \sin \phi_2)^2}} \nonumber \\
\sin \theta & = & \sqrt{1-\cos^2 \theta} \nonumber \\
\sin \phi & = & \sin \phi_2 \nonumber \\
\cos \phi & = & \cos \phi_2
\end{eqnarray}
\item{Block 1 -- (centered about the $+y$-axis; $\pi/4 \le \phi_1 \le 3\pi/4$; $\pi/4 \le \phi_2 \le 3\pi/4$)}
\begin{eqnarray}
\cos \theta & = & \frac{\cos \phi_1 \sin \phi_2}{\sqrt{1-(\cos \phi_1 \cos \phi_2)^2}} \nonumber \\
\sin \theta & = & \sqrt{1-\cos^2 \theta} \nonumber \\
\sin \phi & = & \sin \phi_2 \nonumber \\
\cos \phi & = & \cos \phi_2
\end{eqnarray}
\item{Block 2 -- (centered about the $-x$-axis; $\pi/4 \le \phi_1 \le 3\pi/4$; $3\pi/4 \le \phi_2 \le 5\pi/4$)}
\begin{eqnarray}
\cos \theta & = & \frac{-\cos \phi_1 \cos \phi_2}{\sqrt{1-(\cos \phi_1 \sin \phi_2)^2}} \nonumber \\
\sin \theta & = & \sqrt{1-\cos^2 \theta} \nonumber \\
\sin \phi & = & \sin \phi_2 \nonumber \\
\cos \phi & = & \cos \phi_2
\end{eqnarray}
\item{Block 3 -- (centered about the $-y$-axis; $\pi/4 \le \phi_1 \le 3\pi/4$; $5\pi/4 \le \phi_2 \le 7\pi/4$)}
\begin{eqnarray}
\cos \theta & = & \frac{-\cos \phi_1 \sin \phi_2}{\sqrt{1-(\cos \phi_1 \cos \phi_2)^2}} \nonumber \\
\sin \theta & = & \sqrt{1-\cos^2 \theta} \nonumber \\
\sin \phi & = & \sin \phi_2 \nonumber \\
\cos \phi & = & \cos \phi_2
\end{eqnarray}
\item{Block 4 -- (centered about the $+z$-axis $-\pi/4 \le \phi_1 \le \pi/4$; $-\pi/4 \le \phi_2 \le \pi/4$)}
\begin{eqnarray}
\cos \theta & = & \frac{\cos \phi_1 \cos \phi_2}{\sqrt{1-(\sin \phi_1 \sin \phi_2)^2}} \nonumber \\
\sin \theta & = & \sqrt{1-\cos^2 \theta} \nonumber \\
\sin \phi & = & \frac{\cos \phi_1 \sin \phi_2}{\sin \theta \sqrt{1-(\sin \phi_1 \sin \phi_2)^2}} \nonumber \\
\cos \phi & = & \frac{\sin \phi_1 \cos \phi_2}{\sin \theta \sqrt{1-(\sin \phi_1 \sin \phi_2)^2}} ~.
\end{eqnarray}
\item{Block 5 -- (centered about the $-z$-axis $-\pi/4 \le \phi_1 \le \pi/4$; $-\pi/4 \le \phi_2 \le \pi/4$)}
\begin{eqnarray}
\cos \theta & = & \frac{-\cos \phi_1 \cos \phi_2}{\sqrt{1-(\sin \phi_1 \sin \phi_2)^2}} \nonumber \\
\sin \theta & = & \sqrt{1-\cos^2 \theta} \nonumber \\
\sin \phi & = & \frac{\cos \phi_1 \sin \phi_2}{\sin \theta \sqrt{1-(\sin \phi_1 \sin \phi_2)^2}} \nonumber \\
\cos \phi & = & \frac{-\sin \phi_1 \cos \phi_2}{\sin \theta \sqrt{1-(\sin \phi_1 \sin \phi_2)^2}} ~.
\end{eqnarray}
\end{itemize}

\clearpage
%\bibliographystyle{apj}
%\bibliography{myrefs}

\end{document}